\documentclass[%
reprint,
superscriptaddress,
amsmath,amssymb,
prb,
floatfix]{revtex4-2}

\usepackage{amsthm,wrapfig,pgf,tikz,pgfplots}
\usepackage{graphicx}% Include figure files
\usepackage[permil]{overpic}
\usepackage{dcolumn}% Align table columns on decimal point
\usepackage{bm}% bold math
\usepackage{hyperref}% add hypertext capabilities
\usepackage[normalem]{ulem}

\pgfplotsset{compat=1.15}
\usetikzlibrary{arrows,calc,fadings,decorations.pathreplacing}

\begin{document}

%\preprint{APS/123-QED}

\title{Spin and electronic excitations in $4f$ atomic chains on Au(111) substrates}% 
%\thanks{A footnote to the article title}%

\author{David W. Facemyer}
 \email{df008219@ohio.edu}
\affiliation{Department of Physics and Astronomy and Nanoscale and Quantum Phenomena Institute, Ohio University, Athens, OH 45701}%

\author{Naveen K. Dandu}
\affiliation{Materials Science Division, Argonne National Laboratory, Lemont, IL 60439} 
\affiliation{Chemical Engineering Department, University of Illinois at Chicago, Chicago, IL 60608}%

\author{Alex Taekyung Lee}%
% \email{vsingh2@gitam.edu}
\affiliation{Materials Science Division, Argonne National Laboratory, Lemont, IL 60439} 
\affiliation{Chemical Engineering Department, University of Illinois at Chicago, Chicago, IL 60608}%

\author{Vijay R. Singh}%
% \email{vsingh2@gitam.edu}
\affiliation{Materials Science Division, Argonne National Laboratory, Lemont, IL 60439} 
\affiliation{Chemical Engineering Department, University of Illinois at Chicago, Chicago, IL 60608}%

\author{Anh T. Ngo}
% \email{anhngo@uic.edu}
\affiliation{Materials Science Division, Argonne National Laboratory, Lemont, IL 60439} 
\affiliation{Chemical Engineering Department, University of Illinois at Chicago, Chicago, IL 60608}%

\author{Sergio E. Ulloa}
% \email{ulloa@ohio.edu}
\affiliation{Department of Physics and Astronomy and Nanoscale and Quantum Phenomena Institute, Ohio University, Athens, OH 45701}%

\date{\today}% 

\begin{abstract}
High spin systems, like those that incorporate rare-earth $4f$ elements (REEs), are increasingly relevant in many fields. Although research in such systems is sparse, the large Hilbert spaces they occupy are promising for many applications. In this work, we examine a one-dimensional linear array of europium (Eu) atoms on a Au(111) surface and study their electronic and magnetic excitations. Ab initio calculations using VASP with PBE+U are employed to study the structure.  We find Eu atoms to have a net charge when on gold, consistent with a net magnetic momemt of $\simeq 3.5 \mu_B$.  Examining various spin-projection configurations, we can evaluate first and second neighbor exchange energies in an isotropic Heisenberg model between spin-$\frac{7}{2}$ moments to obtain $J_1 \approx -1.2 \, \mathrm{K}$ and $J_2 \approx 0.2 \, \mathrm{K}$ for the relaxed-chain atomic separation of $a \approx 5$ \AA. These parameters are used to obtain the full spin excitation spectrum of a physically realizable four-atom chain. The large $|J_1|/J_2$ ratio results in a highly degenerate ferromagnetic ground state that is split by a significant easy plane single ion anisotropy of $0.6$ K.\@  Spin-flip excitations are calculated to extract differential conductance profiles as those obtained by scanning tunneling microscopy techniques. We uncover interesting behavior of local spin excitations, especially as we track their dispersion with applied magnetic fields. 
\end{abstract}

%\keywords{Suggested keywords}
\maketitle

%\tableofcontents
%%%%%%%%%%%%%%%%%%%%%%%%%%%%%%%%%%%%%%%%%%%%
\section{\label{Introduction}Introduction}
%-----The importance of spin-chains-----
The study of low dimensional arrays of localized magnetic moments, generally referred to as spin chains, is ubiquitous in physics. The simplicity of such systems resides only in their spatial dimension, however. The classical nature of spin chains is complex enough, but quantum mechanical effects enrich these systems even more. Many models and experiments have allowed us to gather information about the magnetic properties of spin chains, including their ground state and excitations, thermal magnetic properties, and possible phase transitions \cite{choi2019,mikeska2004}.  The last decade has also uncovered new exotic phenomena in atomic chains, such as building unique Kondo configurations \cite{choi2017} or creating Yu-Shiba  \cite{franke2018} or Majorana edge states \cite{yazdani2014,franke2017,wiesendanger2018}  whenever magnetic atoms are placed in proximity to superconducting substrates.  In addition to the  fundamental questions on the behavior of such interesting systems, a great deal of current research is rooted in technological advancement and possible applications, including quantum computing, the transmission of quantum information, and sensing \cite{choi2019,iontrap2019}.

%-----Higher spins are rarely studied-----
Much of this work, however, has been focused on spin-$\frac{1}{2}$ systems, while larger-spin arrays have received much less attention. We consider here spin-$\frac{7}{2}$ magnetic moments as those associated with some lanthanide ($4f$) atoms. The sorts of challenges and promises that come with these systems are easily recognized. Their Hilbert space grows rapidly with increasing site number (chain or cluster size), making them computationally expensive on classical computers. In contrast, these large Hilbert spaces may provide robust computational bases for quantum computation or information storage and a rich state manifold for complex dynamical behavior \cite{lounis2022,dynamics2022}. 

%-----STM interest in spin systems-----
Scanning tunneling microscopy (STM) is a versatile and powerful tool for imaging atomic clusters or molecules on metallic surfaces. STM can further manipulate individual atoms or molecules on a substrate \cite{hla2005,gross2018}, allowing the design and construction of spin chains and clusters with precision, as well as the extraction of single-atom and collective electronic and magnetic properties \cite{Ternes2015,choi2019}. The theoretical work we present here is motivated in part by these extraordinary capabilities, and it aims to provide insight into possible experiments on these deceptively simple structures.

%-----Rare Earth-----
Rare-earth elements are natural physical implementations of large-spin systems. For example, $Gd^{+3}$ ions in $GdRh_2Si_2$ 
are well described experimentally as a pure spin-$\frac{7}{2}$ system \cite{kliemt2017}. Dynamic magnetic properties of core-shell nanowires of such material were studied using an Ising model \cite{ertas2022}, while thermodynamic response functions were obtained in a three-state Blume-Capel model \cite{karimou2021}. Similarly, magnetic properties of $La$-substituted $Gd_{1-x}La_x RhIn_5$ were well characterized using a $\frac{7}{2}$-Heisenberg model for $x \leq 0.5$ \cite{serrano2016}. Beautiful recent experiments exploring lanthanide atoms on thin insulating layers have been able to modify \cite{Ho2017} or measure \cite{Ce2020} the magnetic moment of individual atoms, paving the way for many fundamental and practical questions using such systems. In addition, promising systems of rare-earth molecular complexes have been studied recently \cite{Ajayi2022}.

%-----Relevance to quantum information-----
A wide range of linear array candidates and the means for onsite manipulation also suggest deeper investigations into the realm of quantum information processing. A carefully tuned spin chain with strongly inhomogeneous couplings is predicted to exhibit perfect quantum information transfer \cite{karbach2005}. Likewise, perfect state transmission over many chain sites was demonstrated in both XY and Heisenberg systems with local magnetic fields \cite{christandl2004}. And, quantum state transfer with high fidelity was predicted for large Heisenberg ferromagnetic chains \cite{bose2003}. To accurately substantiate such behavior in future experimental work, we must  accurately obtain model parameters from a rather complex but experimentally realizable system, such as high-spin Eu ions on a Au(111) surface.  

%-----Realistic system with DFT j-couplings-----
To explore such realistic systems, we utilize a combination of theoretical methods. Density functional theory (DFT), an indispensable tool to study the electronic structure of materials, provides us with the characteristic interactions between rare earth atoms such as Eu, deposited on a Au(111) surface. DFT finds a sizable charge and magnetic moment on each atom, compatible with spin-$\frac{7}{2}$, and allows us to extract nearest and next-nearest neighbor exchange energies in a Heisenberg model. Once these parameters are established, we explore a more experimentally realistic structure of an open-ended linear array of Eu atoms on the gold substrate. The low-energy magnetic excitation spectrum is obtained by exact diagonalization, which allows us to analyze the STM spectroscopy for site-specific spin-excitations and their dependence on applied magnetic fields, as well as spatial correlations of the magnetic moments in the chain. The high symmetry of such a system results in low-energy manifolds with interesting magnetic field dependence.  As strong fields cause increasing polarization of an otherwise singlet ferromagnetic ground state, the field-induced transitions are accompanied by characteristic changes in the differential conductance curves.  Such behavior can be easily identified in experiments, providing a direct approach to the energetics of the system and benchmarking of system parameters.  
We also study the role of a possible non-colinear exchange interaction between magnetic moments, and how they would affect the differential conductance curves.  As the different interactions compete with one another in defining the excitation spectrum, they leave signatures that should assist in the full characterization of the overall low-energy dynamics of this rich spin system.   

%%%%%%%%%%%%%%%%%%%%%%%%%%%%%%%%%%%%%%%%%%%%
\section{\label{Methods}Methods}
\subsection{First principles calculations}
Relaxation of a linear chain of Eu atoms on an Au(111) surface was carried out using the Vienna Ab Initio Simulation Package (VASP) \cite{1}. The valence electrons were described in terms of Kohn-Sham orbitals, expanded in a plane-wave basis with an energy cutoff of 500 eV. A $\Gamma$-centered $k$-point mesh of $8 \times 8 \times 1$ is used for all calculations. Atoms were allowed to relax until the net force per atom was less than 0.01 eV/\AA. The supercell was $8 \times 6$, with three gold layers, as well as a vacuum layer thickness of 20 $\mathrm{\AA}$ in the z-direction. We have used PBE+U, with different values for U in Eu: 0, 3 and 7.2 eV. The electronic structure of Eu is qualitatively the same with different U values: Spin-up $f$-bands are fully occupied, while spin-down are fully empty.

\begin{figure}[h]
    \includegraphics[width=8.6cm]{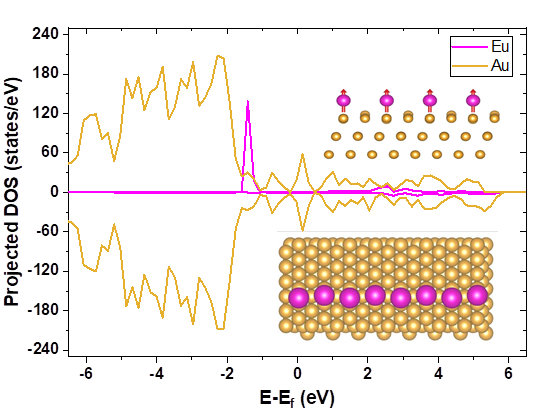}
%        \begin{overpic}[width=8.6cm]{Figs/first spectrum.png}%
    %    \put(520,120){\frame{\includegraphics[width=4cm]{Figs/full spectrum.png}}}%
        %\put(10,620){(1)}%
        %\put(560,360){(2)}%
%        \end{overpic}
        \caption{Projected density of states near the Fermi level of model system for Eu and Au atoms using the PBE+U method with U $=3$ eV (see \cite{Note1} for other U values). Notice spin polarization of the Eu $f$-orbital. Top inset shows side view of Eu atom linear chain on the Au(111) surface. Bottom inset shows top view (two cells) of the system. Supercell has four Eu atoms and a gold slab with three layers; Eu-Eu average distance is 5.01 \AA; Eu-Au distance is 2.8 \AA; interchain separation is 17.3 \AA.\label{DOS}} 
\end{figure}

The simulated system supercell has four Eu-atoms on the gold film with three atomic layers, with adatoms arranged in a way that make the cell repeat periodically, as illustrated in the insets of Fig.\ \ref{DOS}. The bottom gold layer is kept fixed at the bulk lattice parameters, while the rest of the structure is allowed to relax. The electronic projected density of states (PDOS) near the Fermi level of the system indicates that all Eu ions have a well-defined magnetic moment (we find  $\mu \simeq 3.5 \mu_B$) as well as a Bader charge consistent with a 2+ ion -- see Fig.\ \ref{DOS}. Figure S1 (see supplement \cite{Note1}) shows the PDOS of the same system using the different U values. As mentioned, increasing the value of U kept the magnetic moment on Eu unaltered; hereafter, we focus on the U $=7.2$ eV case.

The formation energy of the system with U $=7.2$ eV, calculated from the difference with separate subsystems: $\Delta \mathcal{E}=\mathcal{E}[\mathrm{4Eu \;on \;Au(111)}] - \mathcal{E}[\mathrm{4Eu})] - \mathcal{E}[\mathrm{Au(111}]$, yields $\Delta \mathcal{E}=-12.6$ eV, suggesting the structure is stable. Comparing the energetics of various single Eu atom placement upon the substrate, we find that Eu adsorbed on the bridge of the Au atom gives the more favorable configuration. Table \ref{DFTplacement} compares different locations on gold for Eu atom placement. We should also note that the Eu atoms are not forced to stay in a linear chain. While relaxing the system, adatoms show no propensity to deform the chain. One notices a slight displacement from fully on-bridge locations in the final configuration, likely due to the presence of other Eu atoms.

\begin{table}[h!]
\begin{center}
\begin{tabular}{||l|c|c||} 
 \hline
Eu atom adsorption on  & Relative energy  \\
different Au(111) sites &  [eV]\\ [0.5ex]
 \hline\hline
 \hspace{2ex} Bridge &  0 \\
 \hline
 \hspace{2ex} Hollow HCP & 0.2 \\
 \hline
 \hspace{2ex} On top & 0.3 \\
 \hline
 \hspace{2ex} Hollow FCC & 0.4 \\
 \hline
 \hline
\end{tabular}
\caption{Energetics for different Eu-atom adsorption sites on Au(111) using the PBE+U method with U $=7.2$ eV.}
\label{DFTplacement}
\end{center}
\end{table}

Considering different magnetic configurations for the Eu ions along the chain, we find that the ferromagnetic configuration has the lowest energy, followed by the double-period N\'eel arrangement, while the single-period N\'eel has the highest energy, as shown in Table \ref{DFTresults} -- $U$ and $D$ indicate maximal moment projections $\pm \frac{7}{2}$, respectively.

\begin{table}[h!]
\begin{center}
\begin{tabular}{||c|c|c||} 
 \hline
 Magnetic configuration & Total energy difference  \\
  & $\Delta E$ [meV] \\  [0.5ex]
 \hline\hline
 ${UDUD}$ &  9.92 \\
 \hline
 ${UUDD}$ & 2.89 \\
 \hline
 ${UUUU}$ & 0 \\
 \hline
 \hline
\end{tabular}
\caption{Energy differences per unit cell with respect to the ferromagnetic ($UUUU$) ground state  configuration, for Eu-ion linear chain on Au(111) using the PBE+U method with U $=7.2$ eV.}
\label{DFTresults}
\end{center}
\end{table}

%%%%%%%%%%%%%%%%%%%%%%%%%%%%%%%%%%%%%%%%%%%%
\subsection{Heisenberg model}
In order to extract effective exchange coupling constants, we consider an isotropic Heisenberg Hamiltonian and neglect local magnetic anisotropy \cite{Ternes2015,choi2019}. The Hamiltonian is then 
\begin{equation}\label{FullHam}
    \hat{H} = \sum_{i<j}{\hat{\vec{S}}_i \cdot \mathrm{J}_{ij} \cdot \hat{\vec{S}}_j},
\end{equation}
where $\mathrm{J}_{ij}$ is the full magnetic exchange tensor that couples different pairs of effective moments. We consider first and second-nearest-neighbor exchange coupling, so that the Hamiltonian takes the form
\begin{align}\label{OurHam}
    \hat{H}_{12} &= J_1\sum_{\langle ij \rangle} {\hat{\vec{S}}_i \cdot \hat{\vec{S}}_j} + J_2\sum_{\langle\langle ij \rangle\rangle} {\hat{\vec{S}}_i \cdot \hat{\vec{S}}_j} \nonumber \\
    &= J_1\bigg(\sum_{\langle ij \rangle} {\hat{\vec{S}}_i \cdot \hat{\vec{S}}_j} + \alpha\sum_{\langle\langle ij \rangle\rangle} {\hat{\vec{S}}_i \cdot \hat{\vec{S}}_j}\bigg),
\end{align}
where $\alpha = J_2/J_1$. Recall that $J_{ij} < 0$ corresponds to interacting moments favoring parallel alignment, i.e. ferromagnetic, and $J_{ij} > 0$ favors antiparallel (antiferromagnetic) alignment. One can estimate the coupling constants by evaluating the energy of corresponding configurations in Table \ref{DFTresults}.  Considering the energy for the four-atom supercell, one obtains
\begin{align}
    E_{ F} &= \langle {UUUU} |\hat{H}_{12}| {UUUU} \rangle = 49J_1 \bigg( \alpha + 1 \bigg) \nonumber \\
    E_{N} &= \langle {UDUD} |\hat{H}_{12}| {UDUD} \rangle = 49J_1 \bigg( \alpha - 1 \bigg) \nonumber \\
    E_{dN} &= \langle {UUDD} |\hat{H}_{12}| {UUDD} \rangle = -49J_2 \, .
\end{align}
The energy difference between different configurations are then
\begin{align}
    \Delta E_1 &= E_N - E_F = -98J_1 \nonumber \\
    \Delta E_2 &= E_{dN} - E_F = -49J_1(1+2\alpha). \label{DeltaE}
\end{align}

These equations, together with the corresponding energy differences from Table \ref{DFTresults}, allow the determination of exchange energies, $J_1\simeq -0.101 \, \mathrm{meV \approx -1.2 \, \mathrm{K}}$ and $J_2\simeq +0.021 \, \mathrm{meV \approx 0.2 \, \mathrm{K}}$. (The supplement shows exchange couplings as U values change \cite{Note1}.)

It is important to note that rare-earth ions have been shown to experience appreciable single ion anisotropic (SIA) fields when adsorbed on coinage metal substrates \cite{schuh2012}. 
We have carried out HSE+SOC calculations and find that a single Eu atom on Au(111) has a magnetic easy plane anisotropy $A=0.05$ meV $\approx 0.6$ K, so that the low-energy Hamiltonian for the system is given by
\begin{equation}\label{OurAHam}
    \hat{H}_0 = J_1\sum_{\langle ij \rangle} {\hat{\vec{S}}_i \cdot \hat{\vec{S}}_j} + J_2\sum_{\langle\langle ij \rangle\rangle} {\hat{\vec{S}}_i \cdot \hat{\vec{S}}_j} + A\sum_{i} \big(\hat{S}^{(z)}_i\big) ^2.
\end{equation}

After extracting these parameters from the ab initio calculations, we
employ full diagonalization of the Hamiltonian to explore the spin chain system characteristics, especially its response to external probes.  We use QuSpin \cite{weinberg2017}, and focus here on an open chain with four Eu ions on the gold surface. Notice that the Hilbert space for $N$ sites (magnetic moments) grows as $8^N$ due to the different spin projections, i.e., $m=\{\pm7/2,...,\pm1/2\}$. The $N=4$ chain with open boundary conditions allows us to capitalize on the inversion symmetry of the Hamiltonian and attain reasonable computational times. We can calculate the full magnetic excitation spectrum and corresponding eigenstates, as we discuss below. 

%%%%%%%%%%%%%%%%%%%%%%%%%%%%%%%%%%%%%%%%%%%%
\section{\label{Results}Magnetic excitations and spectroscopy}
%%%%%%%%%%%%%%%%%%%%%%%%%%%%%%%%%%%%%%%%%%%%
\subsection{Magnetic excitation spectrum}

The large ratio $|J_1|/J_2 \simeq 5$ of this system suggests that the system would exhibit strongly  ferromagnetic character throughout the low-lying excitation spectrum, with only a weak frustration induced by $J_2$. In the presence of only $J_1$ (taking $J_2=A=0$), the ground state has maximal spin $S=14$ and corresponding high degeneracy (=29).  A weak $J_2$ does not change the ground state degeneracy but reduces the excitation gap to the $S=13$ manifold from $\simeq 2|J_1|$ to $\simeq 0.8|J_1|$ (see  \footnote{See Supplemental Information LINK}).
\footnote{
It is suggestive to note that the competition between neighbor exchange energies may be related to the indirect exchange mediated by the Au(111) electrons.  The Fermi wavelength for the surface state in Au(111), $\lambda_F \simeq 36 \,\mathrm{\AA}$ \cite{sotthewes2021}, agrees well with the intrachain atomic spacing ($\simeq 5$ \AA) and may be partially responsible for producing the exchange couplings with alternating signs.} 

\begin{figure}[h]
    \begin{overpic}[width=8.6cm]{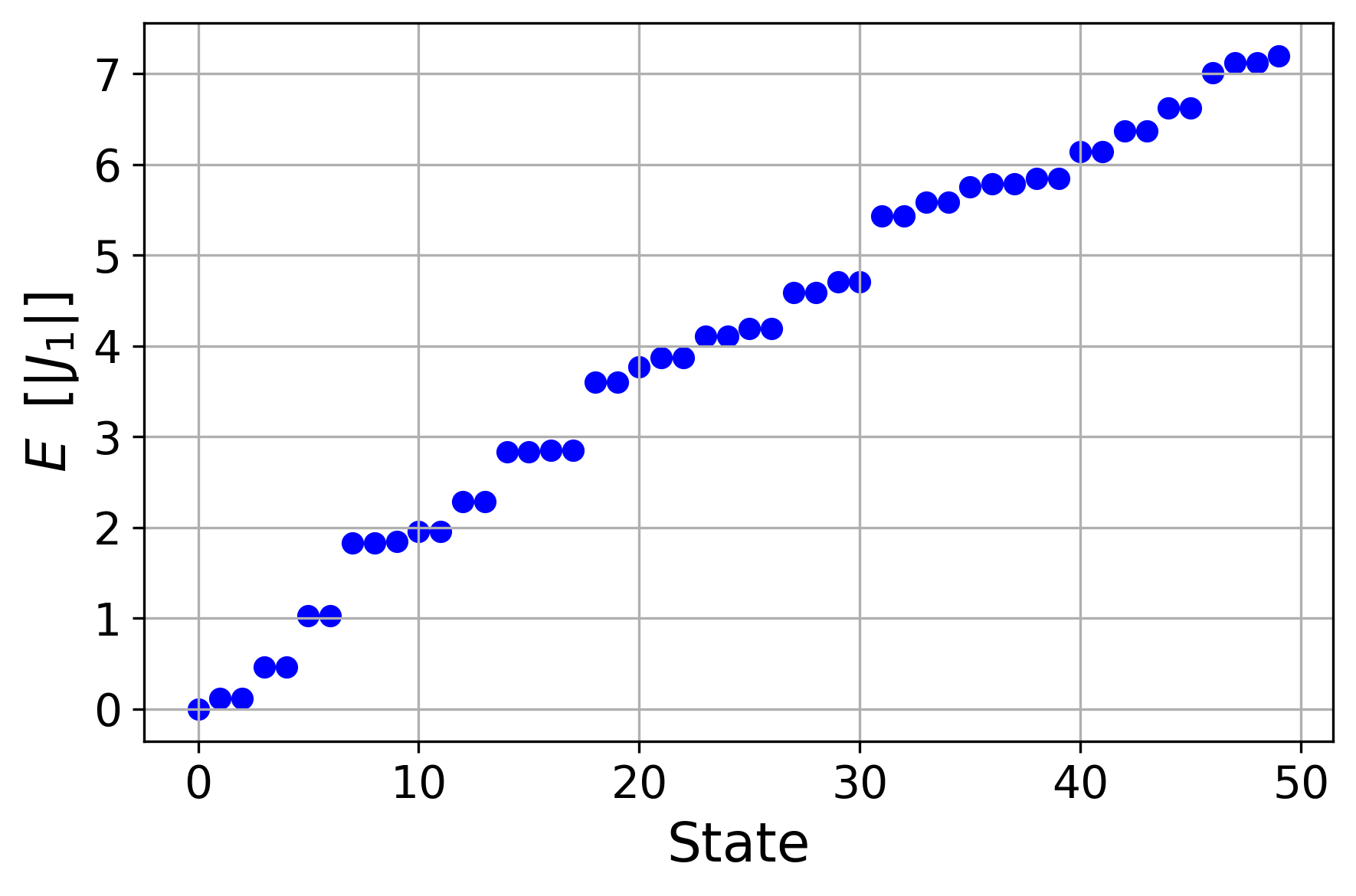}%
    \put(520,120){\frame{\includegraphics[width=4.1cm]{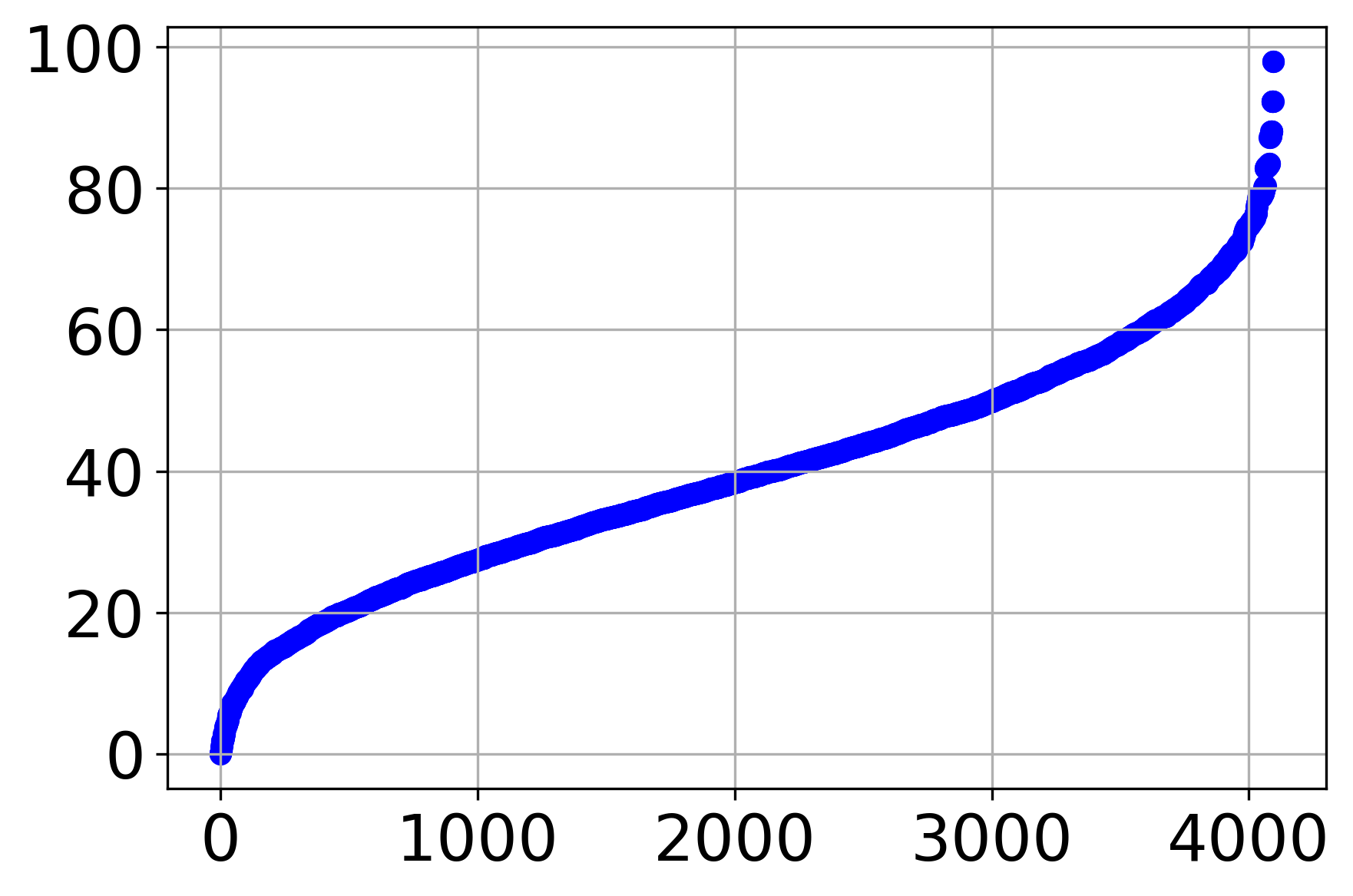}}}%
    %\put(10,620){(1)}%
    %\put(560,360){(2)}%
    \end{overpic}
    \caption{Magnetic excitation spectrum for an open four-site spin chain of $S = 7/2$ magnetic moments with nearest and next-nearest neighbor exchanges, $J_1 = -0.101 \; \mathrm{meV}$ and $J_2 = 0.021 $ meV, as well as an SIA contribution, $A=0.05$ meV. Ground state is non-degenerate. Inset shows full spectral range. Energies in units of $|J_1|$. \label{Spectrum}}
\end{figure}

The presence of SIA in Eq.\ \ref{OurAHam} significantly affects the degeneracy of the $S=14$ ground state, resulting in a non-degenerate `singlet' ($S_z=0$) with ferromagnetic correlations and interesting excitation pattern, as seen in Fig.\ \ref{Spectrum}.
The first excited manifold is a degenerate doublet with $S_z=\pm 1$, followed by doublets with successively larger $S_z$ components, $S_z=\pm 2,\pm 3,\pm 4$. The latter is  closely spaced to another singlet ($S=13, \, S_z=0$) at $\simeq 2|J_1|$ above the ground state. 
The SIA interaction can be seen to produce shifts of the successive $S_z$ doublets by an energy $ \simeq \frac{1}{4}A S_z^2 \simeq \frac{1}{8}|J_1|S_z^2$ -- Fig.\ \ref{Spectrum}.

%%%%%%%%%%%%%%%%%%%%%%%%%%%%%%%%%%%%%%%%%%%%
\subsection{STM spectroscopy}
The rich structure of the magnetic excitation spectrum can be explored using different experimental probes. Ideally suited to carry this out is the differential spectroscopy technique using STM. By increasing bias voltage $V$ while the STM tip is placed at different sites along the chain, one is able to excite different spin configurations that have a local overlap at that site, which are reflected in differential conductance features as peaks in $d^2 I / dV^2$ versus bias \cite{STMSpintorque2010,Ternes2015,choi2019}. We can estimate theoretically what such an experiment will yield by considering the action of the tunneling electron producing spin-raising (or lowering) events on the chain \cite{STMSpintorque2010}.  If the experiment is carried out at low temperatures, the system can be assumed in its ground state $|\Psi_{\rm grnd} \rangle$. As the tunneling electrons go from tip to substrate (or vice versa), they can raise ($S^+$), lower ($S^-$) or leave unchanged ($S^z$) the magnetic moment of the structure they are tunneling through \cite{STMSpintorque2010}. Notice that each tunneling electron can only provide angular momentum changes of $\pm \hbar$ (or 0) on the adsorbed species, so that higher `spin-flips' require repeated electron scatterings before relaxation, and correspondingly higher tunneling currents.

\begin{figure}[h]
    \begin{center}
        \begin{overpic}[width=8.6cm]{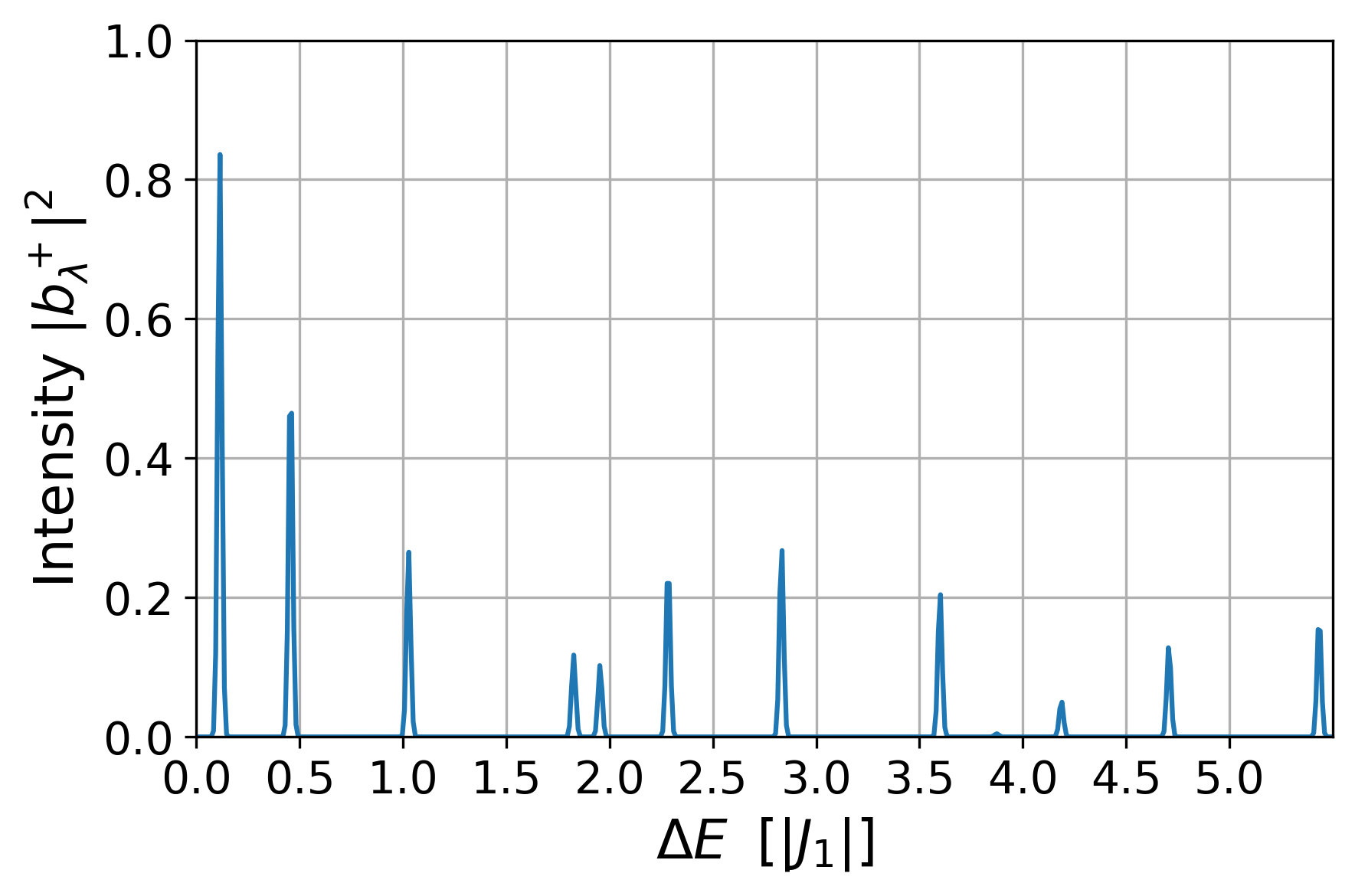}%
        \put(620,390){\frame{\includegraphics[width=3.1cm]{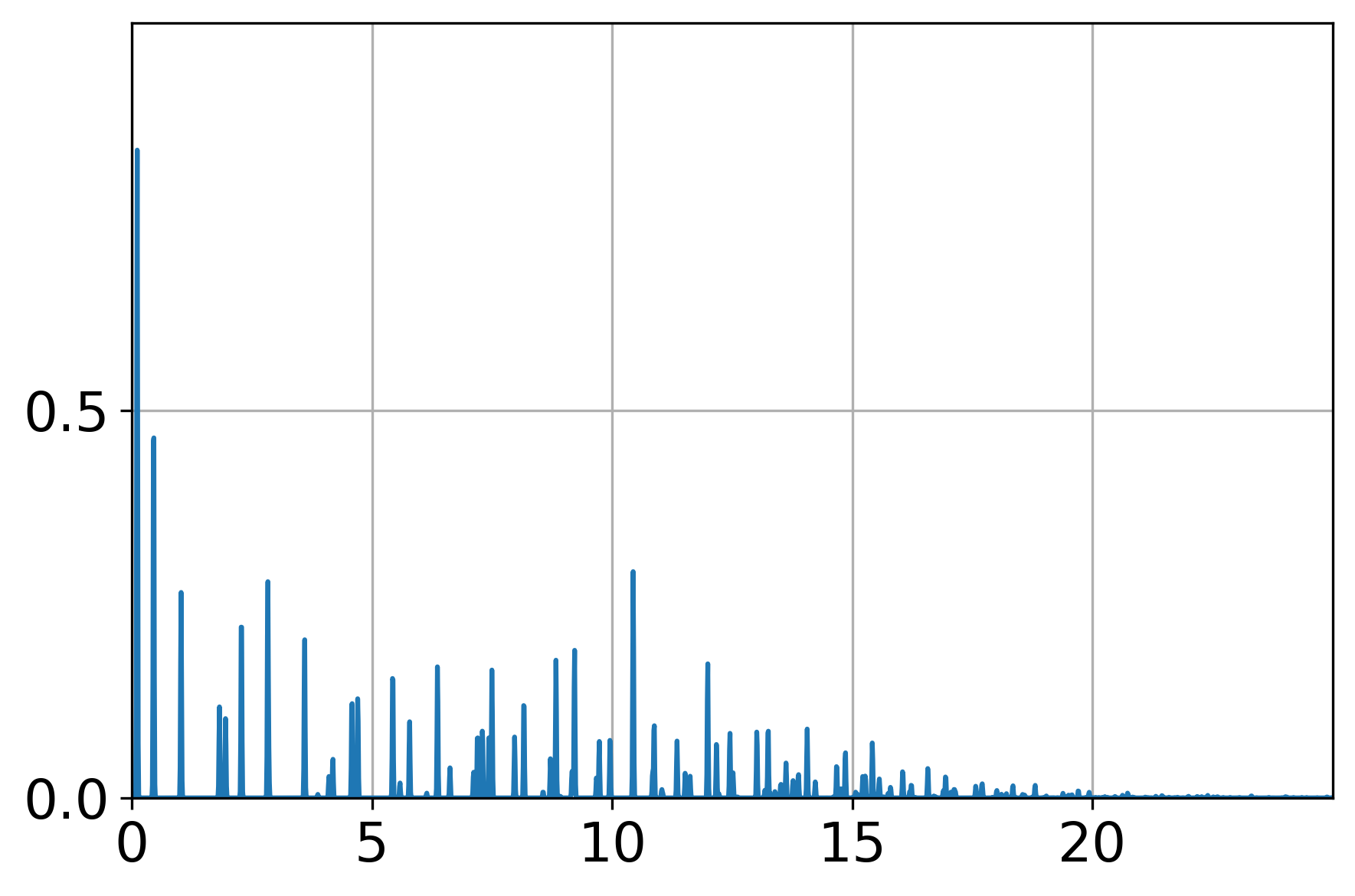}}}%
        \put(250,390){\frame{\includegraphics[width=3.1cm]{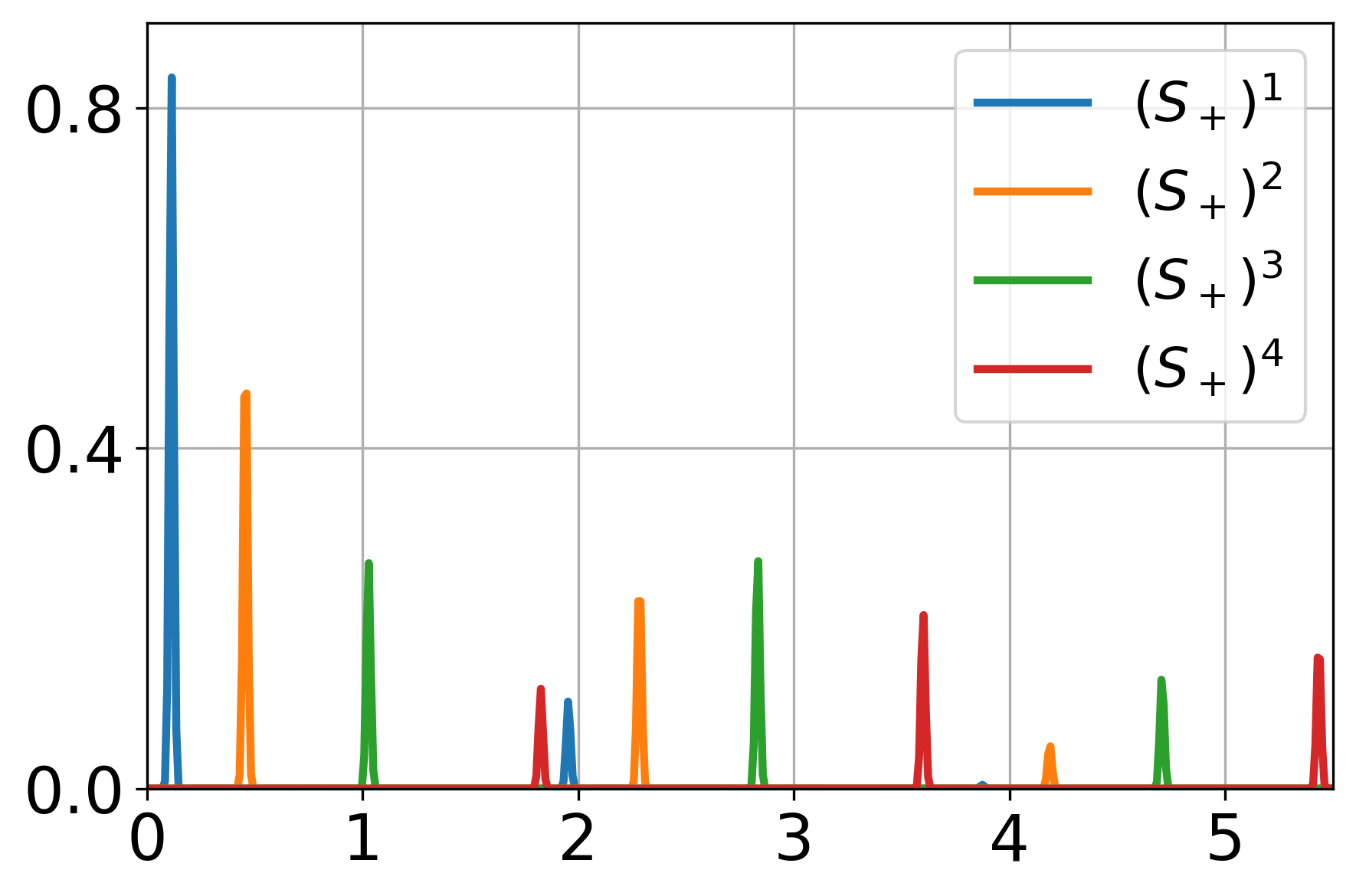}}}%
  %      \put(-20,620){(a)}%
        \put(470,330){\sf SITE 1}%
        \end{overpic} 
    \end{center}
\caption{Excitation profile for repeated spin raising on the first site of the Eu-four-atom chain on its ferromagnetic singlet ground state. Left inset shows contributions to main panel by successive spin 'flips' on site 1 -- higher $n$ appear at higher excitation energy. Right inset shows higher energy region. Weights are displayed using Gaussian broadening with FWHM$\simeq 0.02|J_1|$. All horizontal axes show excitation energy in units of $|J_1|$.  \label{SiteDiffCond}}
\end{figure}

Consider the action of the raising/lower spin operator on the ground state, $(S^{\pm}_{i})^n|\Psi_{\rm grnd}\rangle = |\psi_i^{\pm (n)}\rangle$, which represents a local (on site $i$) magnetic excitation.  This excitation can be seen as composed from different excited eigenstates $|\lambda\rangle$, one can write $|\psi_i^{\pm (n)}\rangle = \sum_{\lambda} b_\lambda^\pm {|\lambda\rangle}$. An example of the effect of successive spin-raising operator actions ($n=1,...,4$) on the first atom of the chain are shown in Fig.\ \ref{SiteDiffCond}, as the resulting weights $|b_\lambda^\pm|^2$ are plotted as a function of the excitation energy $\lambda$. The collection of weights provide an `excitation weight profile' for the system, directly related to the differential conductance curves \cite{toskovic2016}.  This profile provides an estimate of peak-placement and relative intensities for the different transitions \footnote{The visibility of spin excitations will depend on the tunneling current intensity as $\simeq I^n$, so that higher spin `flips' will require higher currents to occur with sufficient amplitude.  Variation of STM set points can provide information on these features.}. 
As the ground state of the chain is a non-degenerate $S_z=0$ state, successive `flips' caused by the tunneling electrons can create excitations along the chain with an overall one-quantum larger or smaller spin. Larger flips involve higher energy, reflecting the strong ferromagnetic correlations of the low-energy excitations.  This provides distinct spectral signatures, as seen by the different color curves in the top left inset of Fig.\ \ref{SiteDiffCond} for different $n$ values, in the case of the spin-raising operator $(S^+_i)^n$. The associated change in current versus bias voltage as different spin-flip channels  open can be easily identified in experiments \cite{heinrich2004,Note3}.

It is interesting to notice that tunneling onto different sites along the chain may change the differential conductance signatures. We find that the first and second sites of the chain have nearly identical spectral yield for low energy, likely again the result of the highly symmetric non-degenerate ground state. However, the amplitudes of various prominent high energy peaks appear much higher when tunneling onto site 1 \cite{Note1}. We emphasize that higher energy excitations would be produced less likely at low tunneling current, resulting in an overall decaying probability envelope for higher energies \cite{Note3}. Also note that the spatial inversion symmetry of the chain ensures the same results for tunneling onto the fourth and third sites, correspondingly. Similar behavior is seen for $(S^-)^n$ on different sites, as expected for an overall singlet ground state.

In what follows, we consider the effect of Zeeman fields on the spectrum of the system, before discussing the weight profile as function of field. 

%%%%%%%%%%%%%%%%%%%%%%%%%%%%%%%%%%%%%%%%%%%%
\subsubsection{Zeeman field effects}
\begin{figure}[h]
    \begin{center}
        \begin{overpic}[width=8.6cm]{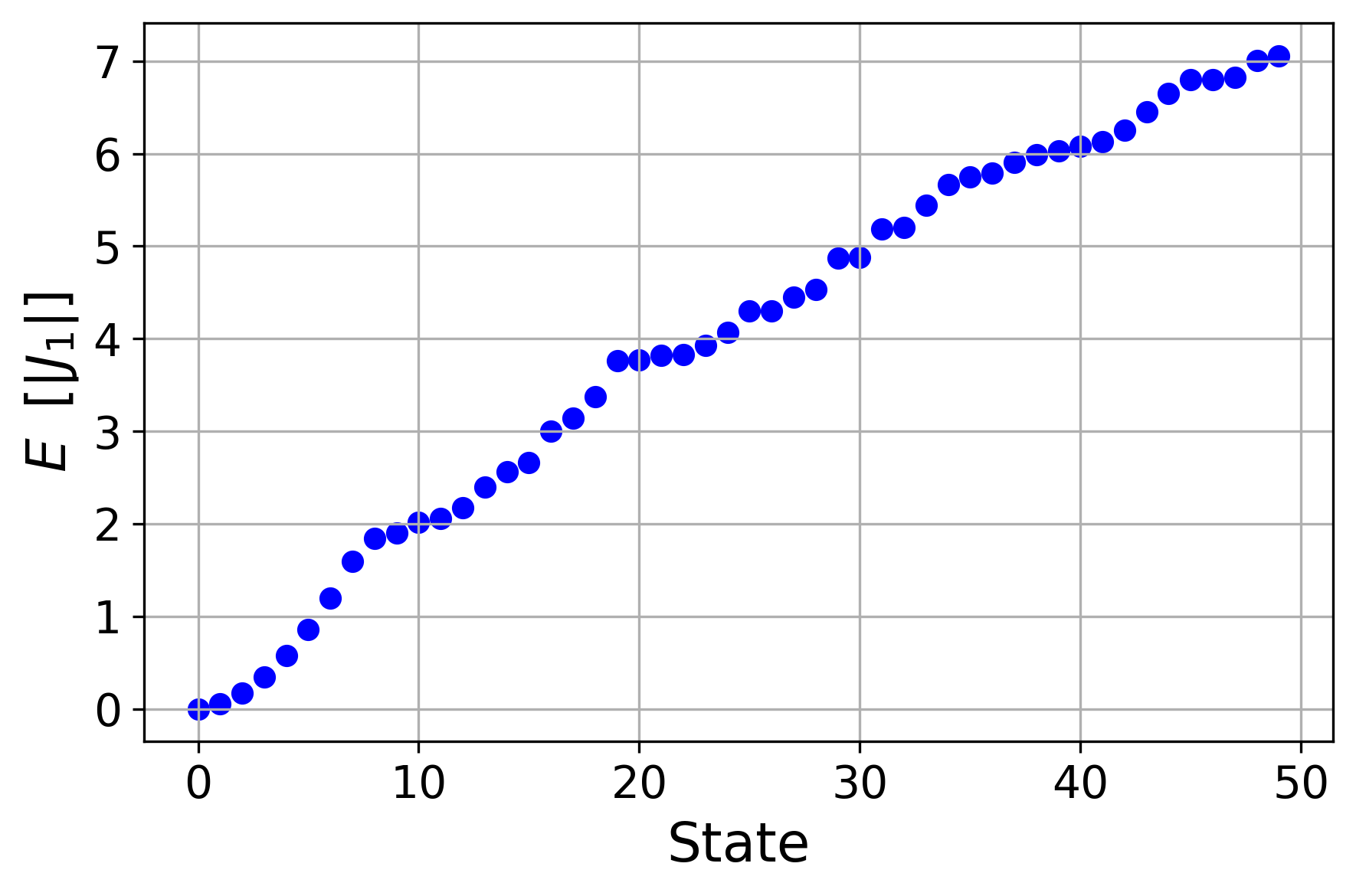}%
        \put(520,120){\frame{\includegraphics[width=4cm]{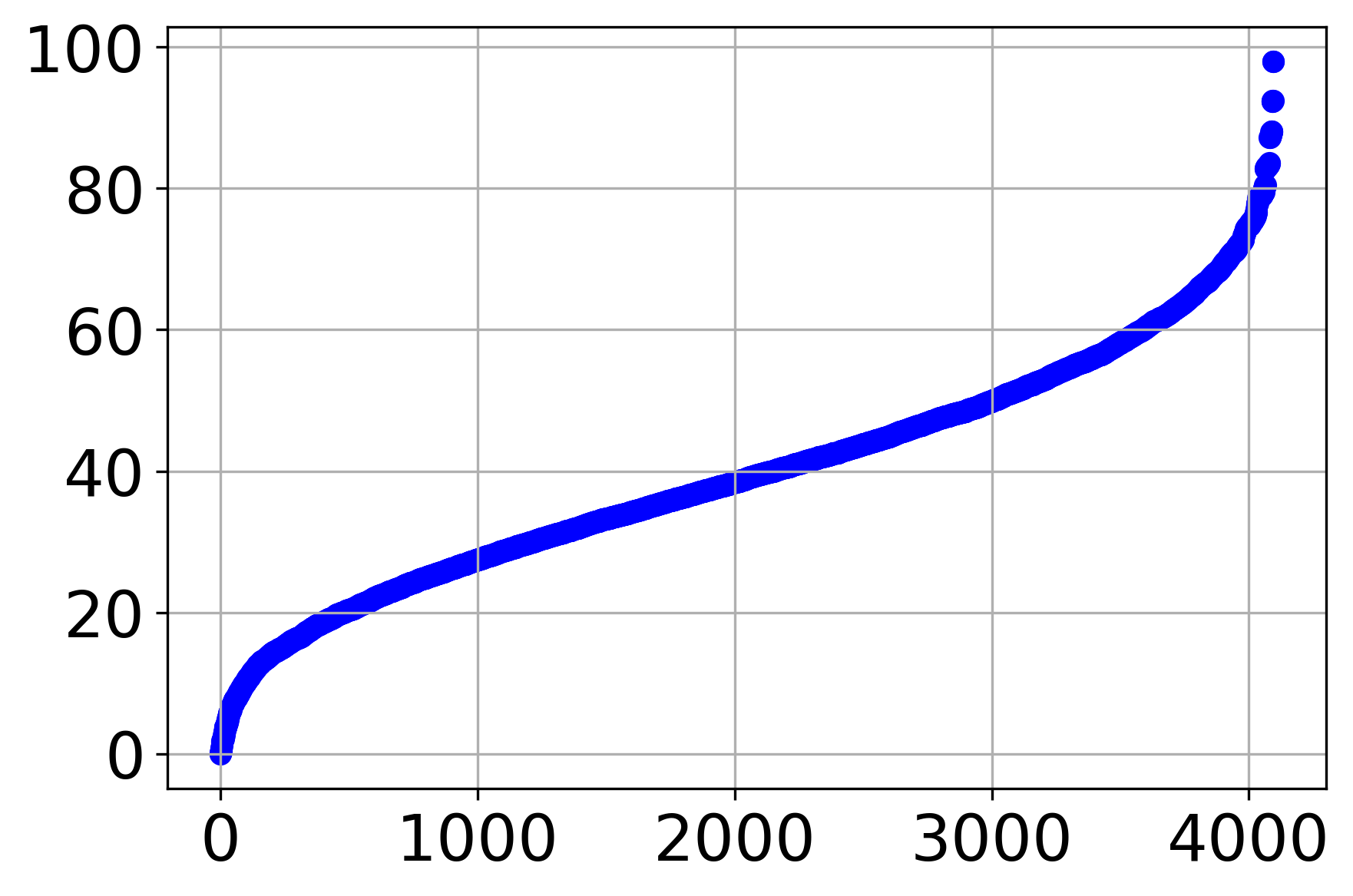}}}%
        %\put(10,620){(1)}%
        %\put(560,360){(2)}%
        \end{overpic} \end{center}
        \caption{Energy excitation spectrum for the four-site spin chain with Zeeman field $h_z=0.1$ T. Notice slight splitting of degenerate multiplets, as compared with those in Fig.\ \ref{Spectrum}. \label{Spectrum_h}}
\end{figure}

\begin{figure}[h]
    \begin{center}
        \begin{overpic}[width=8.6cm]{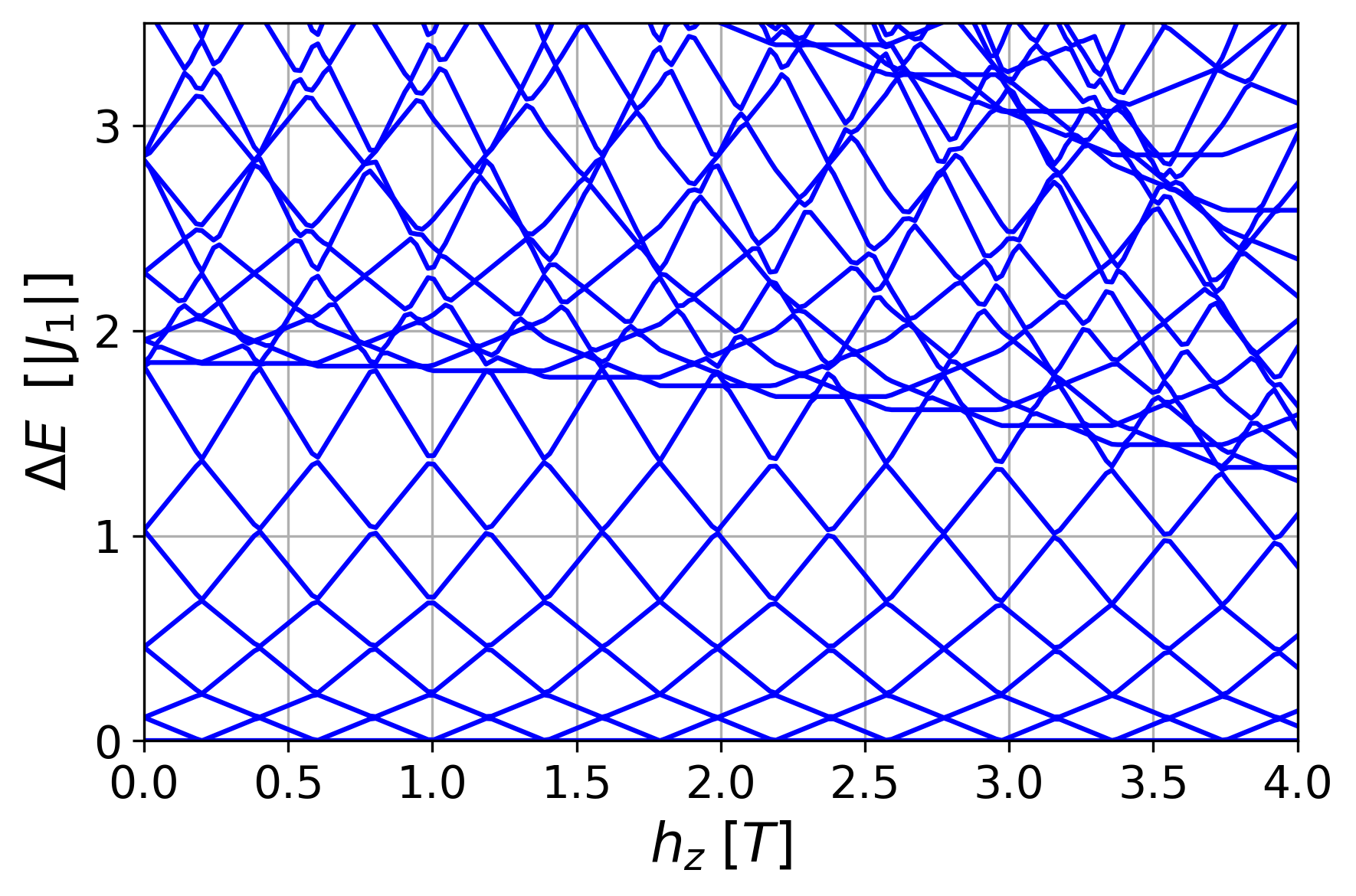}%
        \put(500,350){\frame{\includegraphics[width=4cm]{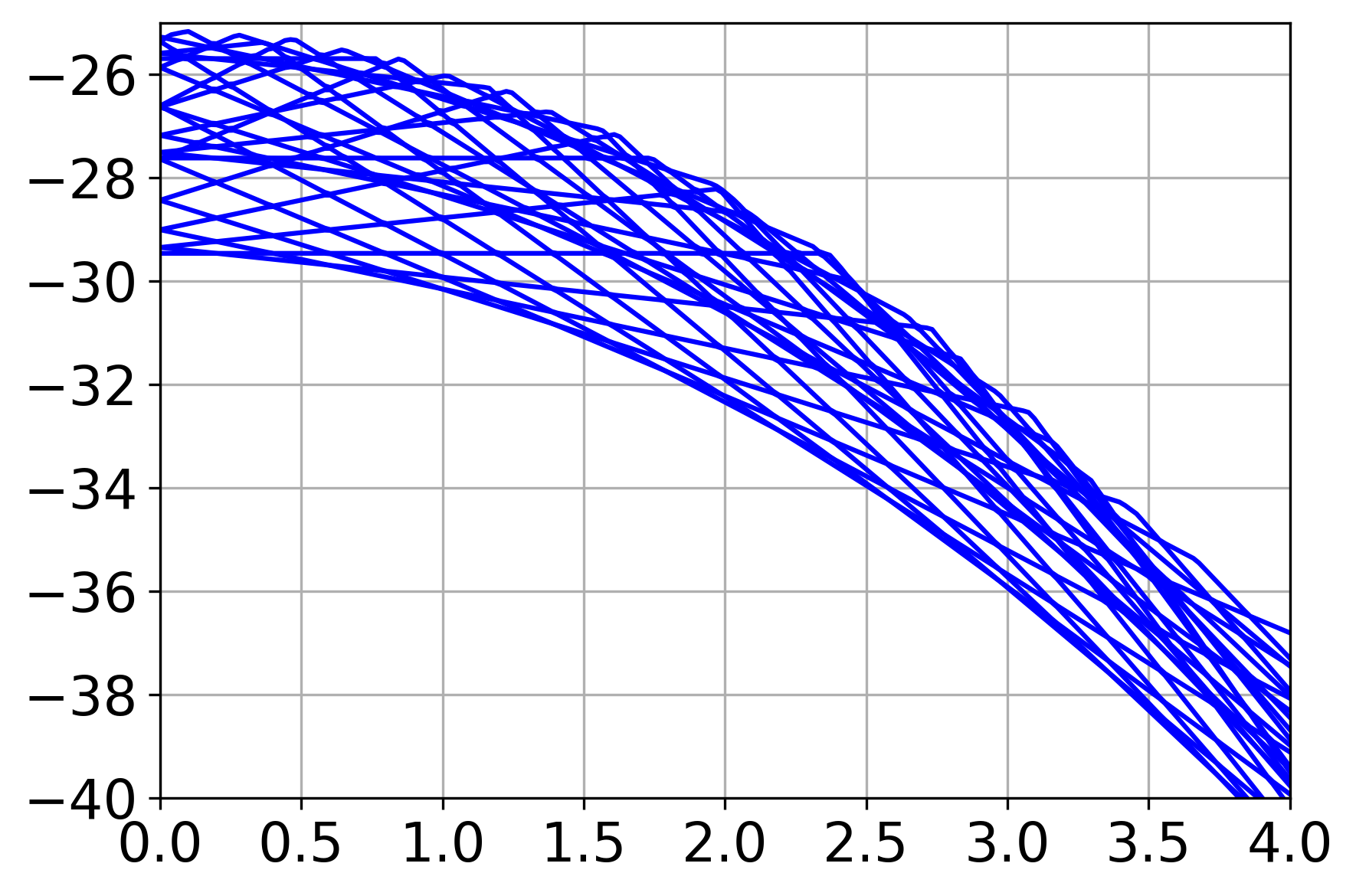}}}%
        %\put(10,620){(1)}%
        %\put(560,360){(2)}%
        \end{overpic} \end{center}
        \caption{ Low-lying eigenstates for increasing applied Zeeman field. The ground state is polarized at increasing field from $S_z=0$ to $-1$, $-2$, etc. at $h_z \simeq 0.2$, 0.6, etc. Main panel shows lowest excitation energies. Inset shows  linear splitting of different states, illustrating their $S_z$ projection, and tangent to a curved lower bound, i.e., the ground state.}  \label{Strong_h}
\end{figure}

Applying magnetic fields during STM spectroscopy experiments provides a powerful tool to obtain further information on the system. A small field of $h_z=0.1 \;\mathrm{T}$ splits degeneracies in the Eu-chain for the low excitation energy manifolds, as shown in Fig.\ \ref{Spectrum_h}. The field splittings are linear, as the perturbing Hamiltonian is $\hat{H}_Z = h_z \hat{S}_z$, associated with the corresponding $S_z$ total-spin projection of each state, so that the full Hamiltonian is
\begin{equation}\label{OurZHam}
    \hat{H} = \hat{H}_0 + h_z\sum_{i} \hat{S}^{(z)}_i \, ,
\end{equation}
with $\hat{H}_0$ in Eq.\ \ref{OurAHam}.
The field dependence of the low-energy excitation spectrum is displayed in Fig.\ \ref{Strong_h}. The ground state of the chain is polarized from $S_z=0$ to $-1$ at $h_z \gtrsim 0.2 \, \mathrm{T}$, with decreasing $S_z$ projection as $h_z$ increases (to $S_z=-2$ at $\approx 0.6$ T, etc.).

\begin{figure}[h]
    \begin{center}
        \begin{overpic}[width=8.6cm]{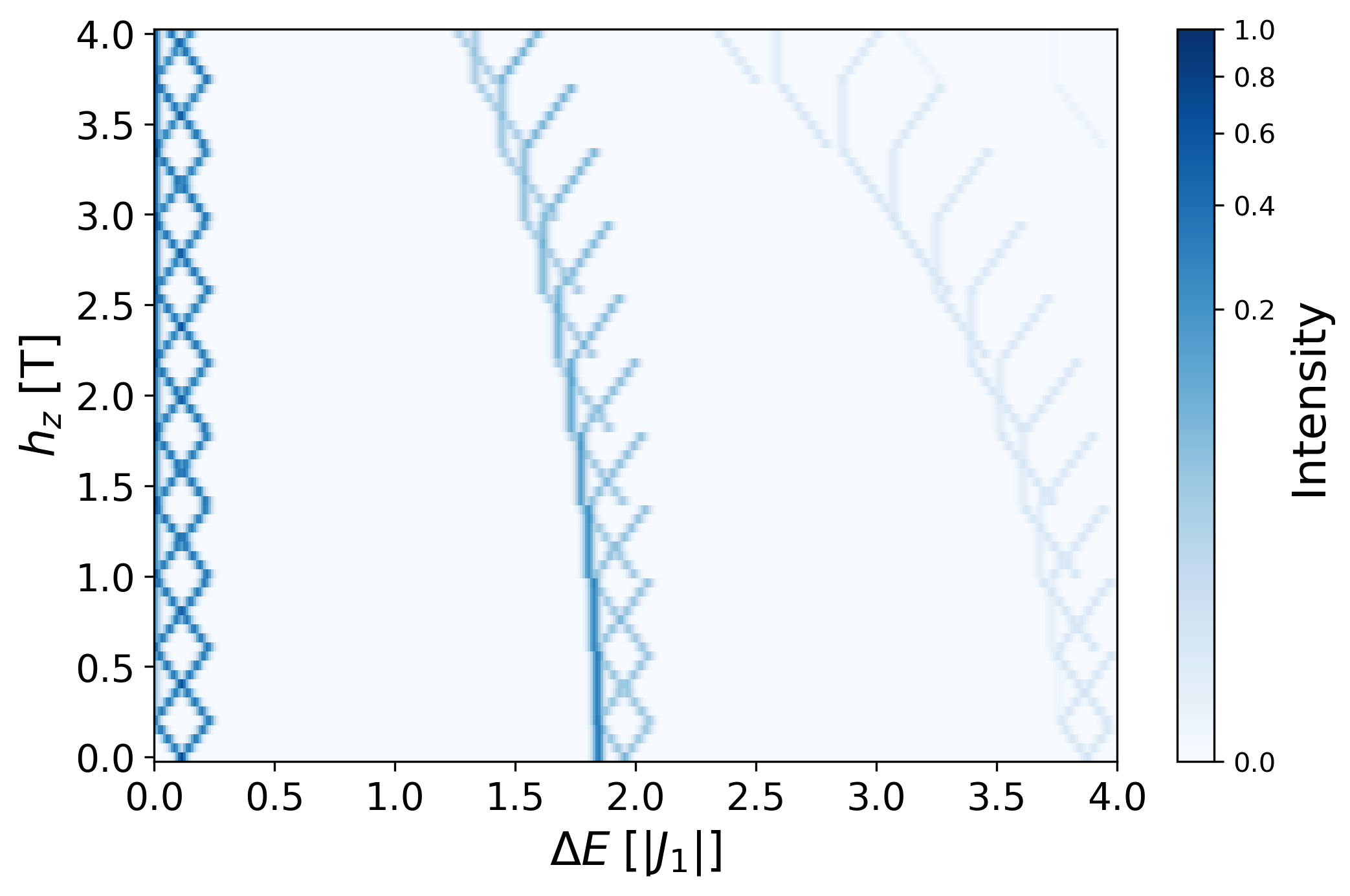}%
        \put(200,580){(a)}%
        \end{overpic}
        \begin{overpic}[width=8.6cm]{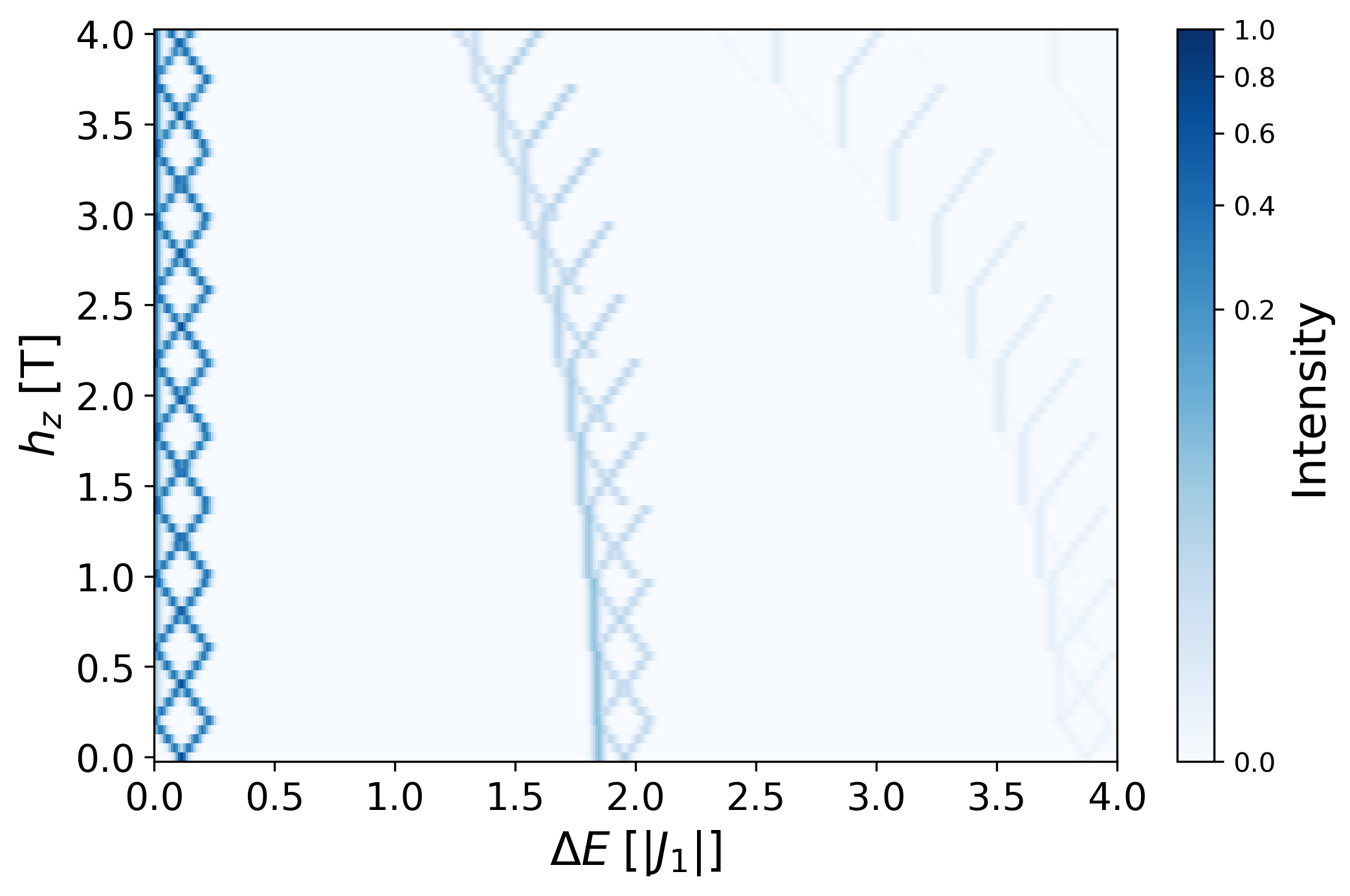}%
        \put(200,580){(b)}%
        \end{overpic}
    \end{center}
\caption{Heatmap of excitation weight profile ($\simeq$ differential conductance signatures) for increasing Zeeman field strength and corresponding energy (bias).  We show $I = 2|b^z|^2 + |b^+|^2 + |b^-|^2$ to account for all different scattering processes for tunneling electrons as the energy/bias increases. Results for tunneling through (a) first and (b) second site. Tunneling through the second site produces weaker intensities, especially at higher energy excitations and stronger fields, as seen in (b).} \label{DiffCond_heatmap}
\end{figure}

To explore the role of the Zeeman field on the differential conductance features, we calculate the excitation weight profile on the first and second sites of the chain as a function of field. As seen in the spectrum, the Zeeman coupling results in clear shifts with increasing field, as well as successive transitions to ever increasing $|S_z|$ projection reflected well in the excitation weight profile.  Figure \ref{DiffCond_heatmap}a and b, show profiles for single-scattering/tunneling events for site 1 and 2, respectively. We see that the low-energy profiles for the different site excitations are rather similar, although much weaker peaks at higher energy/bias are seen on site 2--this reflects the different environment of the two sites, which pin site 2 more strongly onto ferromagnetic order with the chain.  For completeness, excitation weight profiles for higher-order scattering events are shown in \cite{Note1,Note3}.

%%%%%%%%%%%%%%%%%%%%%%%%%%%%%%%%%%%%%%%%%%%%
\subsubsection{Non-collinear exchange effects}
\begin{figure}[h]
    \begin{center}
        \begin{overpic}[width=8.6cm]{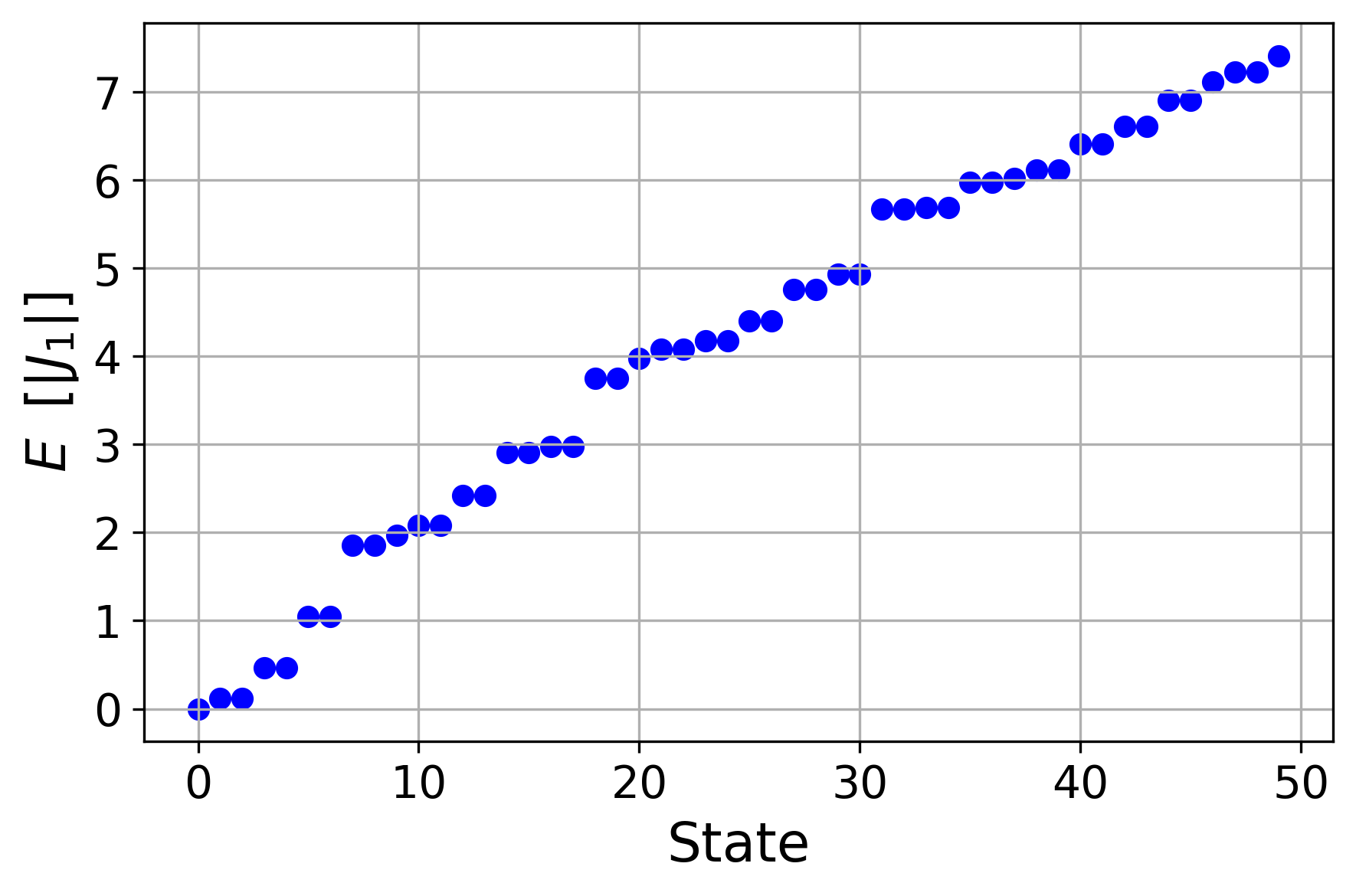}%
        \put(520,110){\frame{\includegraphics[width=4cm]{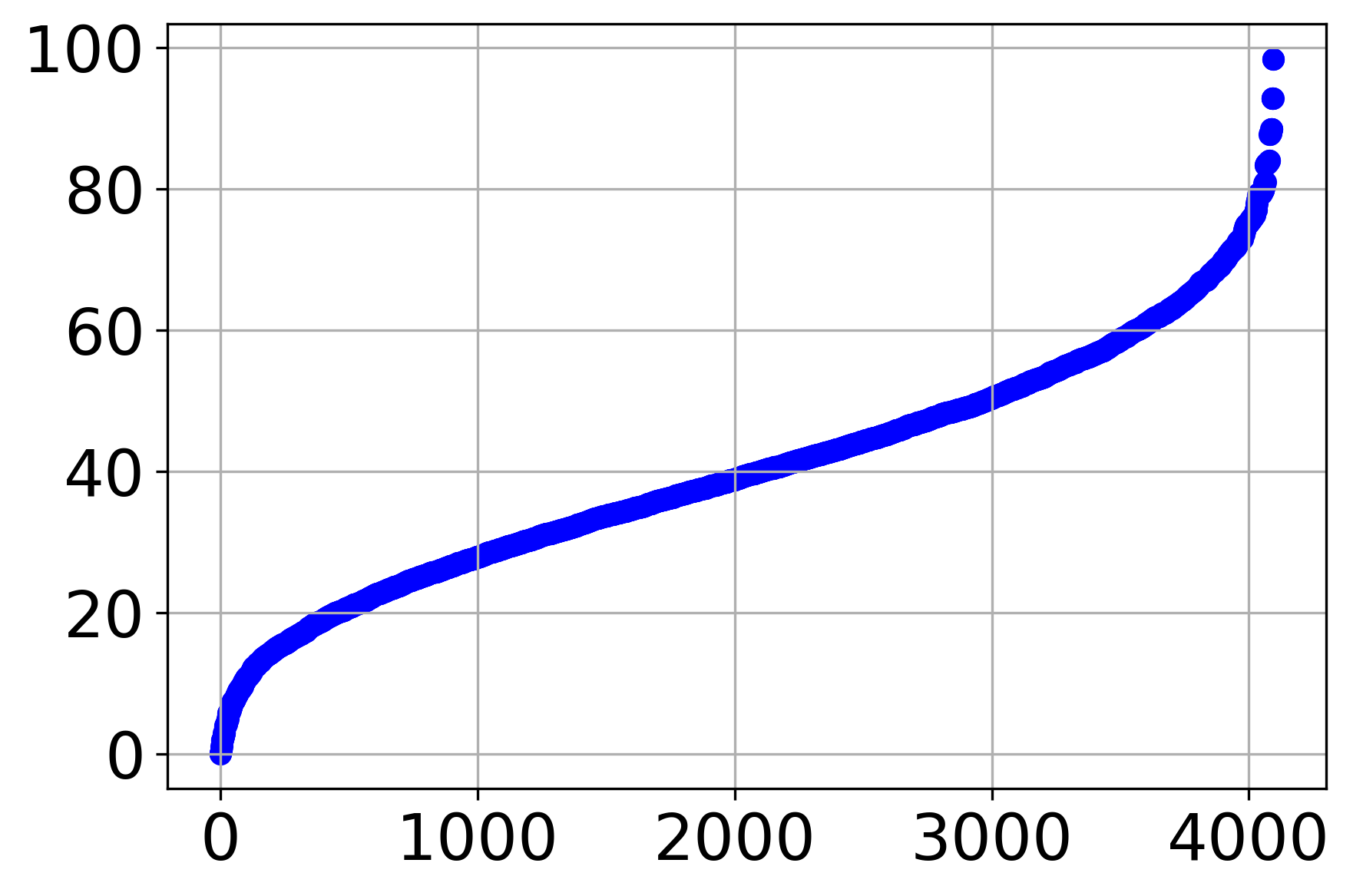}}}%
        %\put(10,620){(1)}%
        %\put(560,360){(2)}%
        \end{overpic}    \end{center}
        \caption{Low energy and (inset) full excitation spectrum for the four-site spin chain with $h_z=0$ and $D_z=0.1| J_1|$. Compare with Fig.\ \ref{Spectrum}. \label{Spectrum_d}}
\end{figure}

\begin{figure}[h]
    \begin{center}
        \begin{overpic}[width=8.6cm]{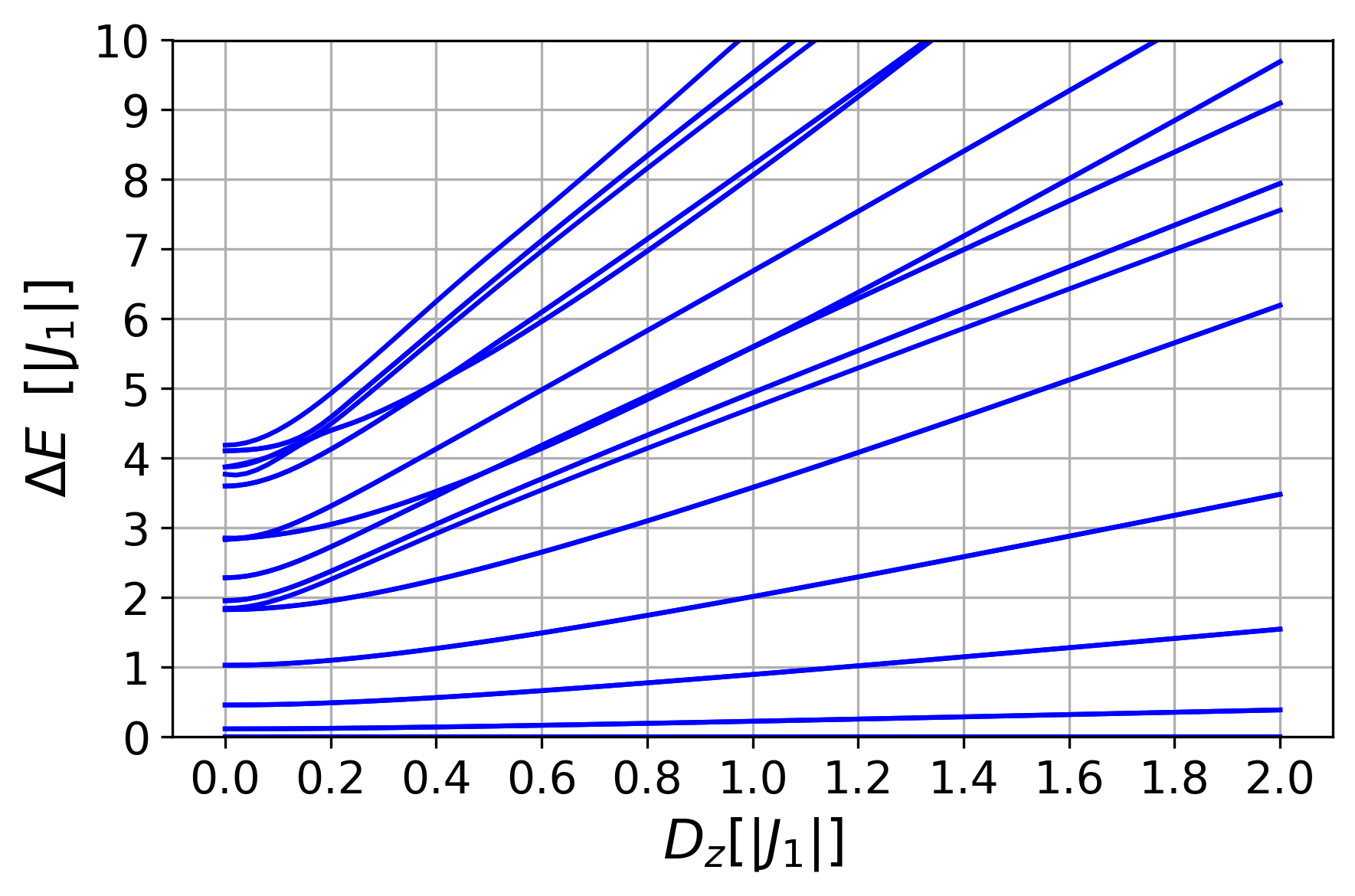}%
        \put(520,110){\frame{\includegraphics[width=4cm]{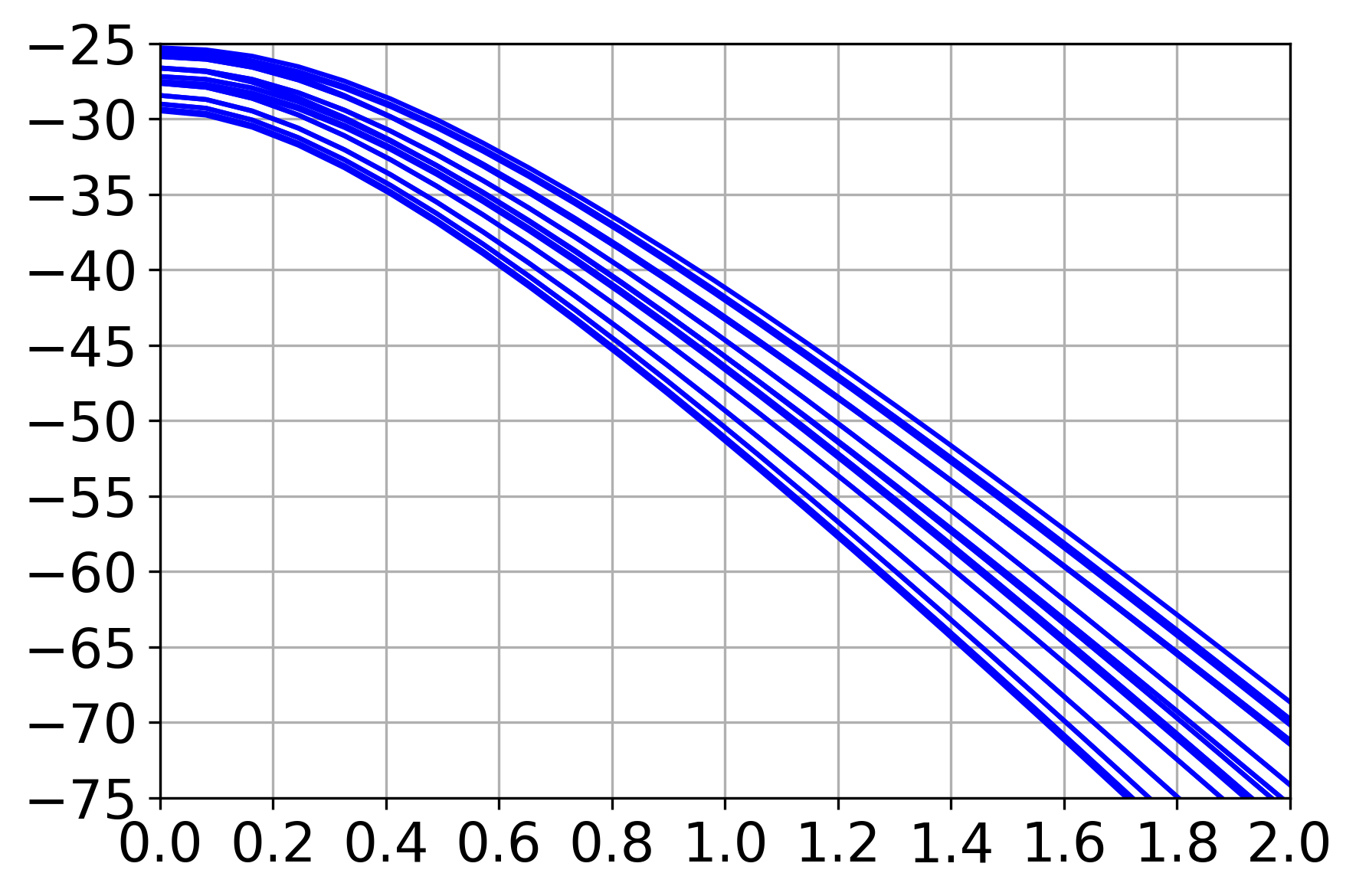}}}%
        %\put(10,620){(1)}%
        %\put(560,360){(2)}%
        \end{overpic} \end{center}
        \caption{Main panel shows excitations for Eu-chain as $D_z$ increases. Inset shows corresponding low-lying excitations, all for $h_z=0$. \label{Varying_d}}
\end{figure}

It is reasonable to expect a Dzyaloshinskii-Moriya interaction (DMI) may also appear in the spin system due to the strong Rashba spin-orbit effect measured on the gold surface \cite{hoesch2004}. This interaction results in the antisymmetric Hamiltonian $H_{DMI}=\sum_{ij} \vec{D}\cdot \hat{\vec{S}}_i \times \hat{\vec{S}}_j$,  that can qualitatively change the ground state configuration as well as the rest of the excitation spectrum \cite{D1958,M1960,VernekChiral}.  To assess its effect on STM spectroscopy, we consider the likely dominant $z$-component of the DMI vector $\vec{D}$. The full Hamiltonian then is
\begin{align}\label{OurDHam}
    \hat{H} &= \hat{H}_{\rm Eq.\ \ref{OurZHam}} + \frac{iD_z}{2}\sum_{\langle ij \rangle}\big(\hat{S}_{i}^{+} \hat{S}_{j}^{-} - \hat{S}_{j}^{+}\hat{S}_{i}^{-}\big).
\end{align}
Figure \ref{Spectrum_d} displays the resulting spectrum with $D_z=0.1 |J_1| = 0.01$ meV, showing that the ground state remains non-degenerate, while the excitation doublets are split more strongly the larger their $|S_z|$ values -- notice that the $\pm S_z$ degeneracies in the spectrum remain. Figure \ref{Varying_d} shows how the excitation spectrum changes for larger $D_z$ values, shifting all energies down as $D_z$ increases, splitting degenerate manifolds, and producing some level crossings at high energy in the excitation spectrum with $D_z$ (main panel).

%

%\vspace{1em}

\begin{figure}[h]
    \begin{center}
        \begin{overpic}[width=8.6cm]{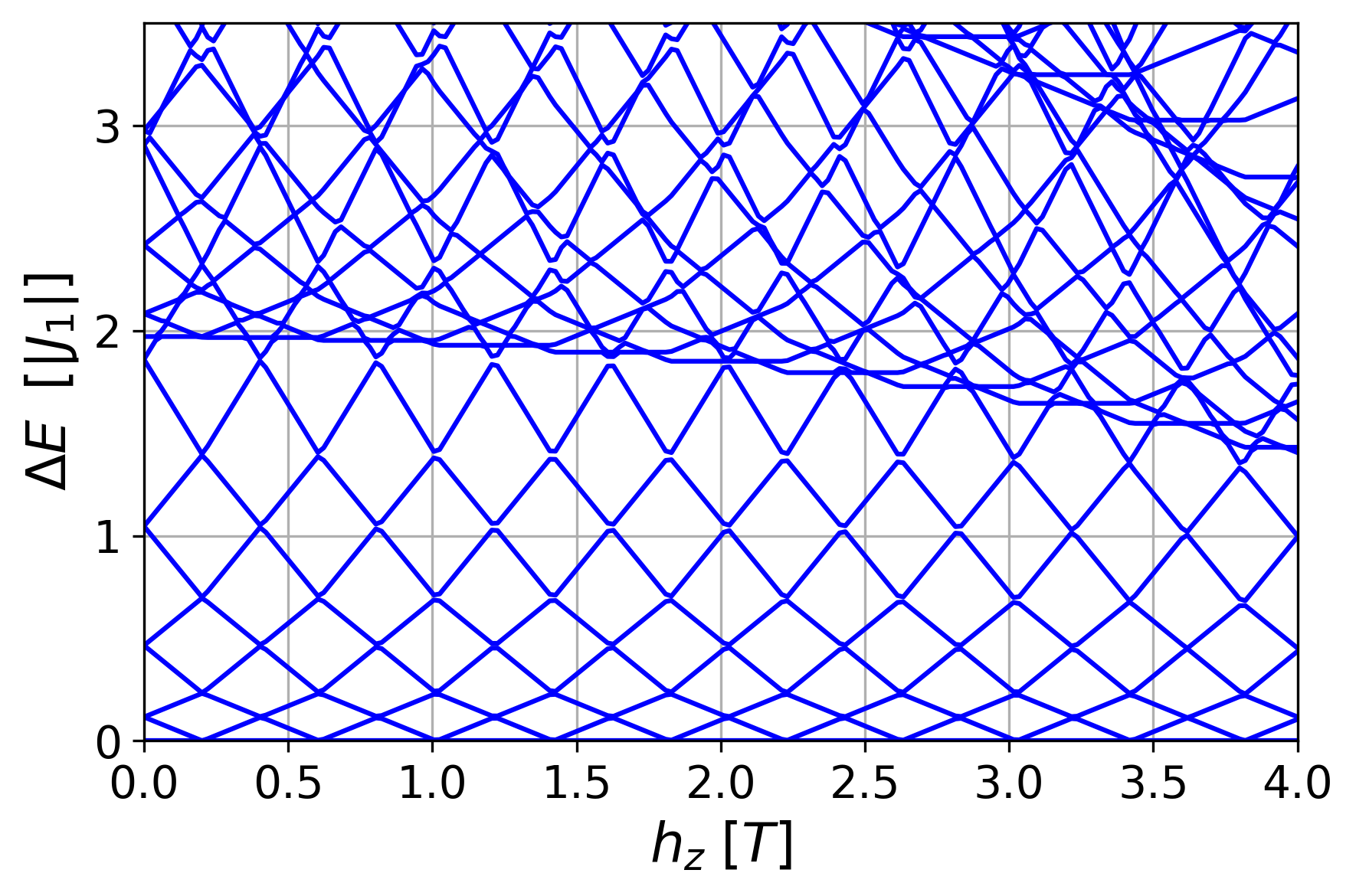}%
        %\put(520,100){\frame{\includegraphics[width=4cm]{Figs/full spectrum h=01 d=01.png}}}%
        %\put(10,620){(1)}%
        %\put(560,360){(2)}%
        \end{overpic}     \end{center}
        \caption{Low-lying eigenstates for increasing applied Zeeman field and $D_z=0.1 |J_1|$ for the system of Fig.\ \ref{Spectrum}. \label{Strong_h_with_d}}
\end{figure}

\begin{figure}[h]
    \begin{center}
        \begin{overpic}[width=8.6cm]{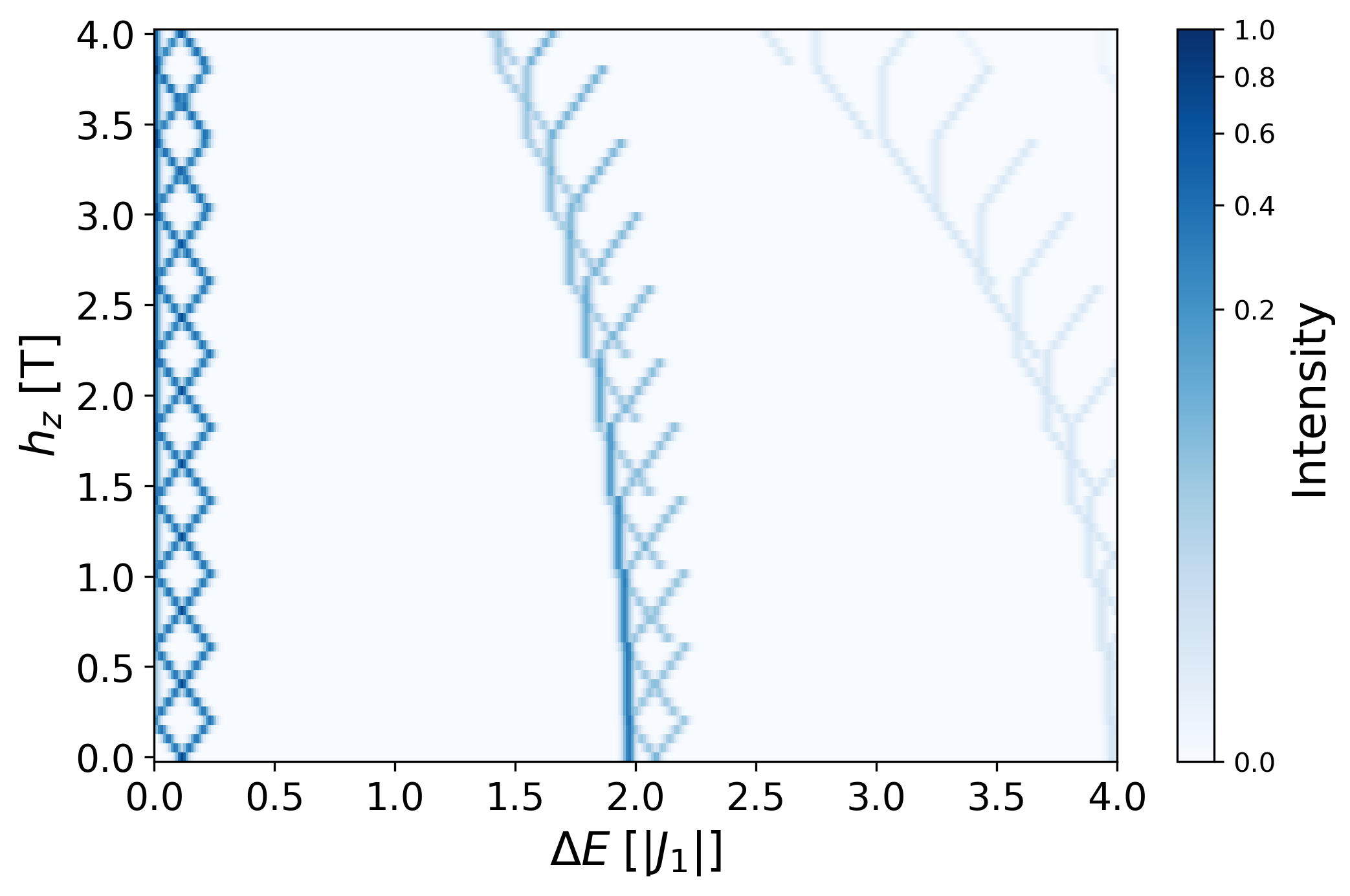}%
    %    \put(4,650){(a)}%
        \end{overpic}
    \end{center}
\caption{Heatmap of excitation weight profile for increasing Zeeman field and corresponding energy (bias) for the system with $D_z=0.1|J_1|$ in Fig.\ \ref{Strong_h_with_d}.  We plot $I = 2|b^z|^2 + |b^+|^2 + |b^-|^2$ to account for  different scattering processes for tunneling electrons. Results for tunneling through first site. \label{DiffCond_D_heatmap}} 
\end{figure}

We calculate the excitation weight profile as a function of Zeeman field for the system in Fig.\ \ref{Spectrum_d}, with field dependence shown in Fig.\ \ref{Strong_h_with_d}. Notice that the successively increasing polarization of the ground state appears here as well at lower Zeeman field strength.  However, the $S_z=0$ state in the second excitation manifold shifts from $\simeq 1.8 |J_1|$ (Fig.\ \ref{Strong_h}) to about $2|J_1|$ for this $D_z$ value. These shifts produce a general rightward bias (upward in energy) shift in the prominent peaks of the weight profiles in Fig.\ \ref{DiffCond_D_heatmap}. Without DMI, these peaks begin at zero-field near $\Delta E \simeq 1.8|J_1|$ and $\simeq 3.7|J_1|$; with DMI, they shift to start $\Delta E \simeq 2$ and $\simeq 4|J_1|$, and the single-tunneling events at higher energy are somewhat stronger. In this case (not shown), tunneling through site 2 produces essentially the same differential conductance profile, indicating that the DMI interaction has reduced the site asymmetry, as one would expect.  

%%%%%%%%%%%%%%%%%%%%%%%%%%%%%%%%%%%%%%%%%%%%
\section{\label{Conclusion}Conclusions}
The advent of atomic-site manipulation via STM has heralded a new era of materials science and engineering. One can not only design quantum systems, but precisely assemble them and measure their electronic and magnetic characteristics at the level of a single atom or molecular entity. Spin chains, with a great history of deep insights into the physics of interacting systems, garner ever increasing interest as they naturally interface with growing fields of quantum information transmission, computing, and sensing. Our proposal to assemble large-spin systems to highlight their rich behavior has focused on a four-atom chain of effective Eu$^{2+}$ ions on a Au(111) surface.  First principles calculations of realistic systems allow us to extract magnetic excitation spectra and resulting differential conductance profiles with site-specific excitations.  Experimental realization of this system would provide us with the interesting opportunity of not only benchmarking our model and specific exchange parameters, but would be the next step in exploring quantum information transfer down long chains. 

%%%%%%%%%%%%%%%%%%%%%%%%%%%%%%%%%%%%%%%%%%%%

\begin{acknowledgments}
We thank S.-W. Hla and E. Masson for helpful discussions and acknowledge financial support from the U.S. Department of Energy, Office of Science, Office of Basic Energy Sciences, Materials Science and Engineering Division.
\end{acknowledgments}

%\appendix

%\section{Appendixes}

%\section{A little more on appendixes}

%\subsection{\label{app:subsec}A subsection in an appendix}

% The \nocite command causes all entries in a bibliography to be printed out
% whether or not they are actually referenced in the text. This is appropriate
% for the sample file to show the different styles of references, but authors
% most likely will not want to use it.
%\nocite{*}

\bibliography{spin}% Produces the bibliography via BibTeX.

\end{document}

\preprint{APS/123-QED}

\title{Supplement Information for \\Spin and electronic excitations in $4f$ atomic chains on Au(111) substrates}% Force line breaks with \\

\author{David W. Facemyer}
% \email{df008219@ohio.edu}
\affiliation{Department of Physics and Astronomy and Nanoscale and Quantum Phenomena Institute, Ohio University, Athens, OH 45701}%

\author{Naveen K. Dandu}
\affiliation{Materials Science Division, Argonne National Laboratory, Lemont, IL 60439} 
\affiliation{Chemical Engineering Department, University of Illinois at Chicago, Chicago, IL 60608}%

\author{Alex Taekyung Lee}
\affiliation{Materials Science Division, Argonne National Laboratory, Lemont, IL 60439} 
\affiliation{Chemical Engineering Department, University of Illinois at Chicago, Chicago, IL 60608}%

\author{Vijay R. Singh}%
\affiliation{Materials Science Division, Argonne National Laboratory, Lemont, IL 60439} 
\affiliation{Chemical Engineering Department, University of Illinois at Chicago, Chicago, IL 60608}%

\author{Anh T. Ngo}
% \email{anhngo@uic.edu}
\affiliation{Materials Science Division, Argonne National Laboratory, Lemont, IL 60439} 
\affiliation{Chemical Engineering Department, University of Illinois at Chicago, Chicago, IL 60608}%

\author{Sergio E. Ulloa}
% \email{ulloa@ohio.edu}
\affiliation{Department of Physics and Astronomy and Nanoscale and Quantum Phenomena Institute, Ohio University, Athens, OH 45701}%

%\date{\today}% 

\maketitle

\section{PBE+U Results}
\begin{figure}[h]
        \begin{overpic}[width=8.6cm]{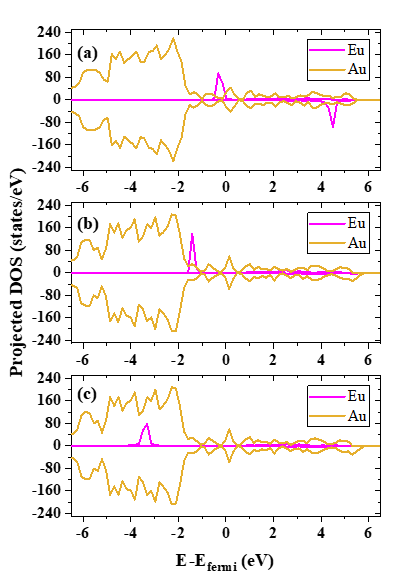}%
        \end{overpic}
        \caption{Projected density of states near the Fermi level for Eu and Au atoms of the model system in the main text, using the PBE+U method. a) U$=0$, b) U$=3$ eV, and c) U$=7.2$ eV. \label{S1}}
    \end{figure}

Figure \ref{S1} represents PDOS of the system, as described in Fig.\ 1 of the main text, using the PBE+U method. We have used three different values of U: 0, 3 and 7.2 eV.  Although the Eu $f$-orbital shows some change/shifts, we find these changes do not affect the value of magnetic moment of each Eu ion.  

Similarly, we explore the energetics of different magnetic configurations for the different U values.  The corresponding exchange parameters are shown in Table \ref{DFTmethods}.
\begin{table}[h!]
\begin{center}
\begin{tabular}{||l|c|c||} 
 \hline
PBE + U & $J_{nn}$ [meV] & $J_{nnn}$ [meV] \\ [0.5ex]
 \hline\hline
 U = 0 eV &  0.07 & 0.05 \\
 \hline
 U = 3 eV & -0.08 & 0.03 \\
 \hline
 U = 7.2 eV & -0.10 & 0.02 \\
 \hline
 \hline
\end{tabular}
\caption{Exchange coupling strengths for different U values.}
\label{DFTmethods}
\end{center}
\end{table}

\section{Heisenberg Hamiltonian}
\subsection{Next-nearest neighbor interaction}
\begin{figure}[h]
    \begin{overpic}[width=8.6cm]{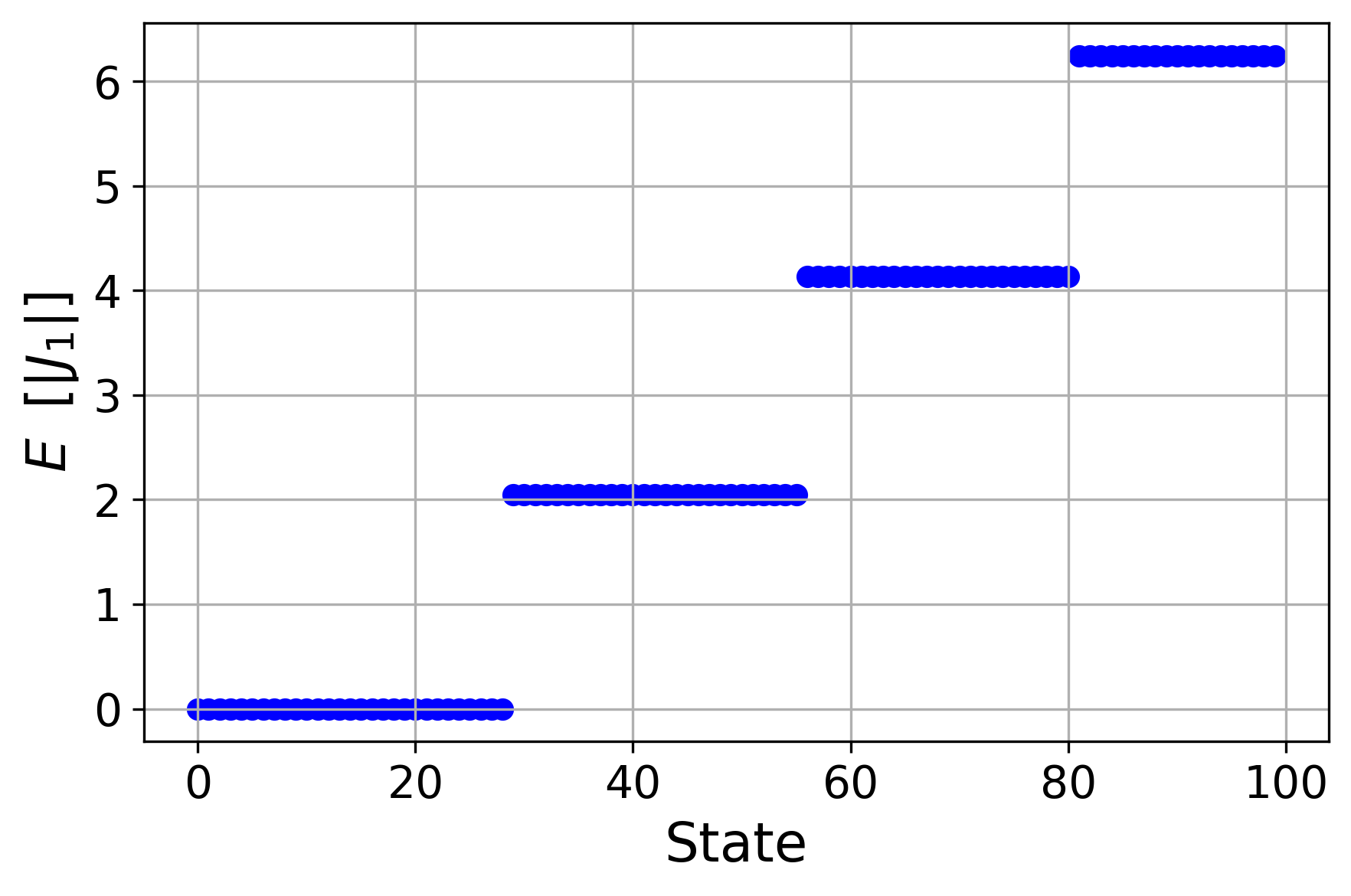}%
       \put(520,120){\frame{\includegraphics[width=4cm]{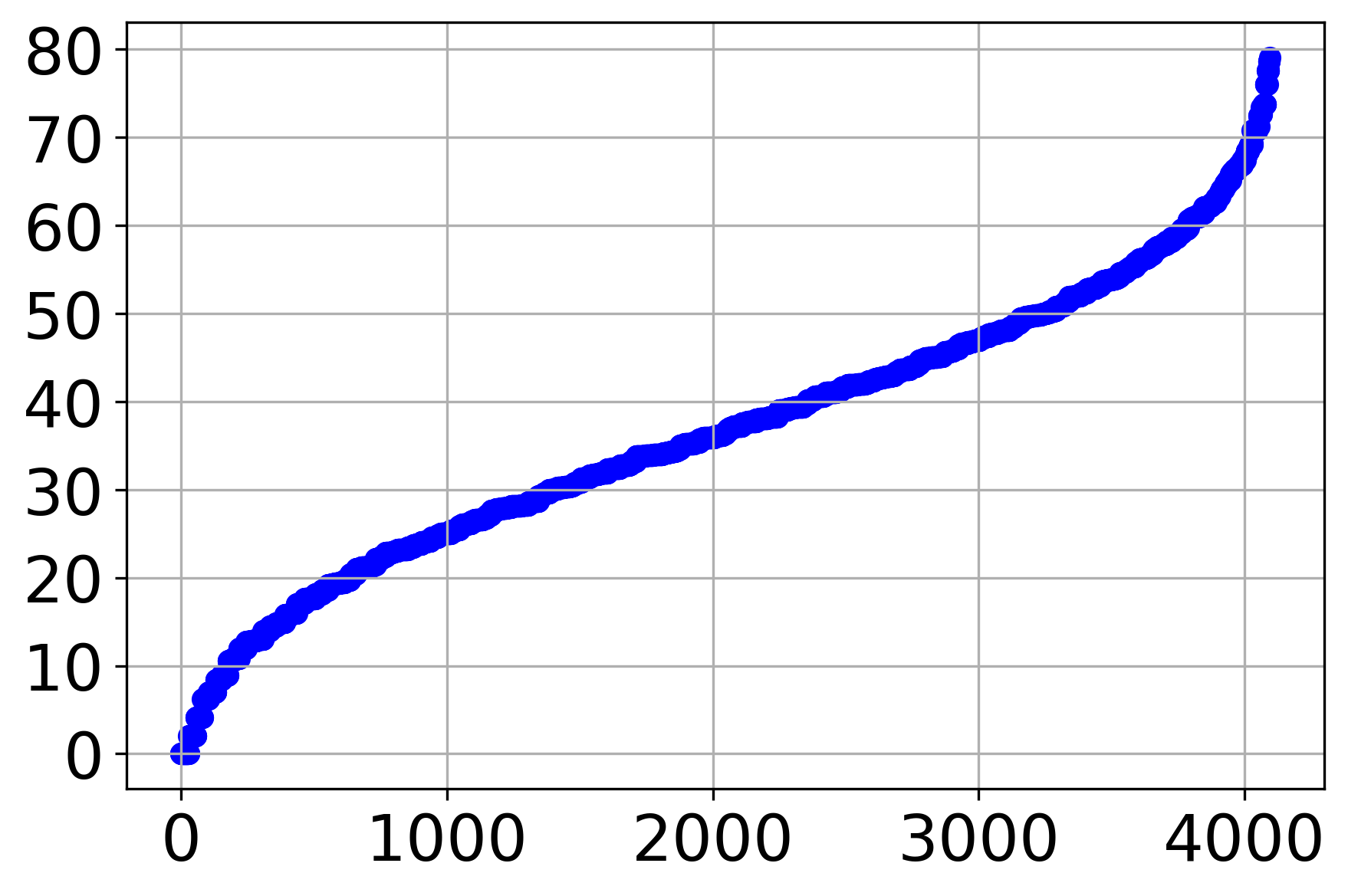}}}%
        %\put(10,620){(1)}%
        %\put(560,360){(2)}%
    \end{overpic}
    \caption{Spectrum of four-site linear chain with $A=J_2=0$.\label{SpectrumNoSIANoJ2}}
\end{figure}

The dominance of the nearest-neighbor exchange coupling characterizes the system. Due to the small strength of the second-neighbor coupling, we could consider the case when $J_2=0$ and analyze its effect as a perturbation. If we also ignore single-ion anisotropy (and set $A=0$), the spectrum appears as in Fig.\ \ref{SpectrumNoSIANoJ2}.  Notice the high degeneracy of each manifold, associated with different values of the total spin, where the maximum value is $S=14$, and corresponds to the lowest-energy manifold of 29 states.  At $\Delta E \simeq 2|J_1|$ sits the $S=13$ manifold (with 27 states), followed by $S=12$ at $\Delta E \simeq 4|J_1|$ (25 states), etc.  

\begin{figure}[]
    \begin{overpic}[width=8.6cm]{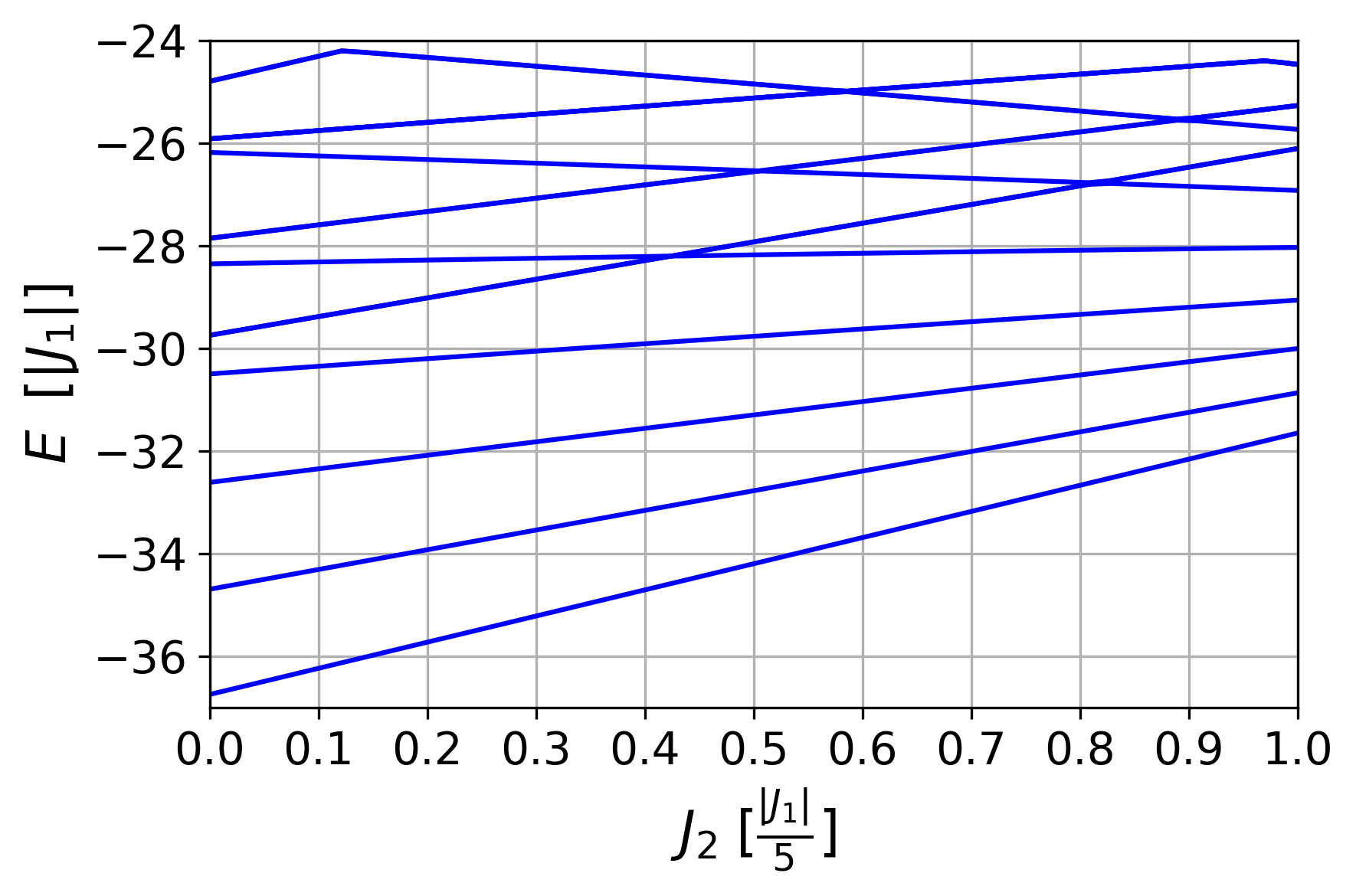}%
    \end{overpic}
    \caption{Spectrum of four-site linear chain with increasing $J_2$ antiferromagnetic coupling, while $J_1$ is fixed and without SIA ($A=0$). The ground state manifold remains 29-fold degenerate, with linear energy increase with $J_2$ over this range. \label{j2_pert_noA}}
\end{figure}
\begin{figure}[h]
    \begin{overpic}[width=8.6cm]{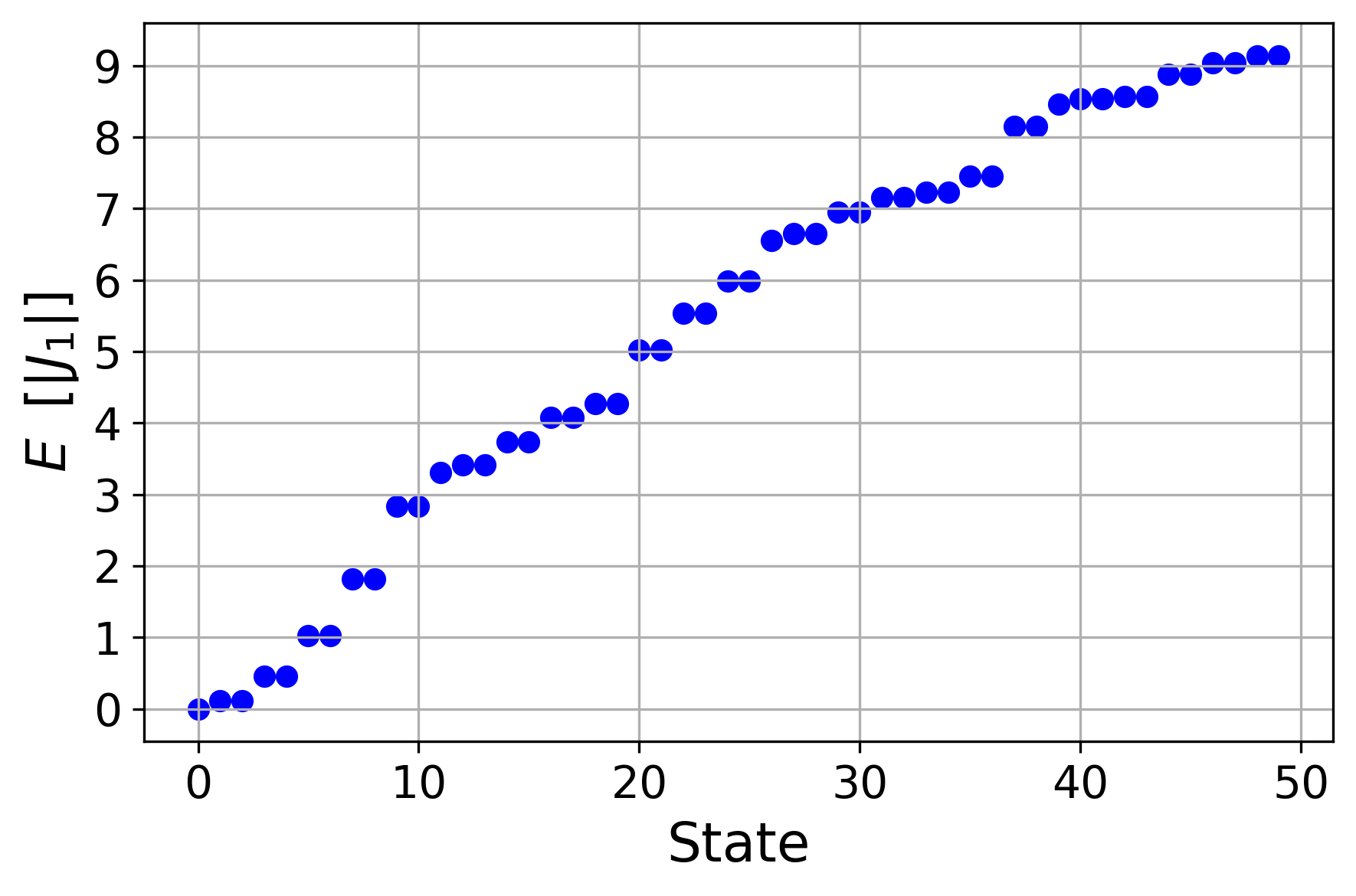}%
       \put(520,120){\frame{\includegraphics[width=4cm]{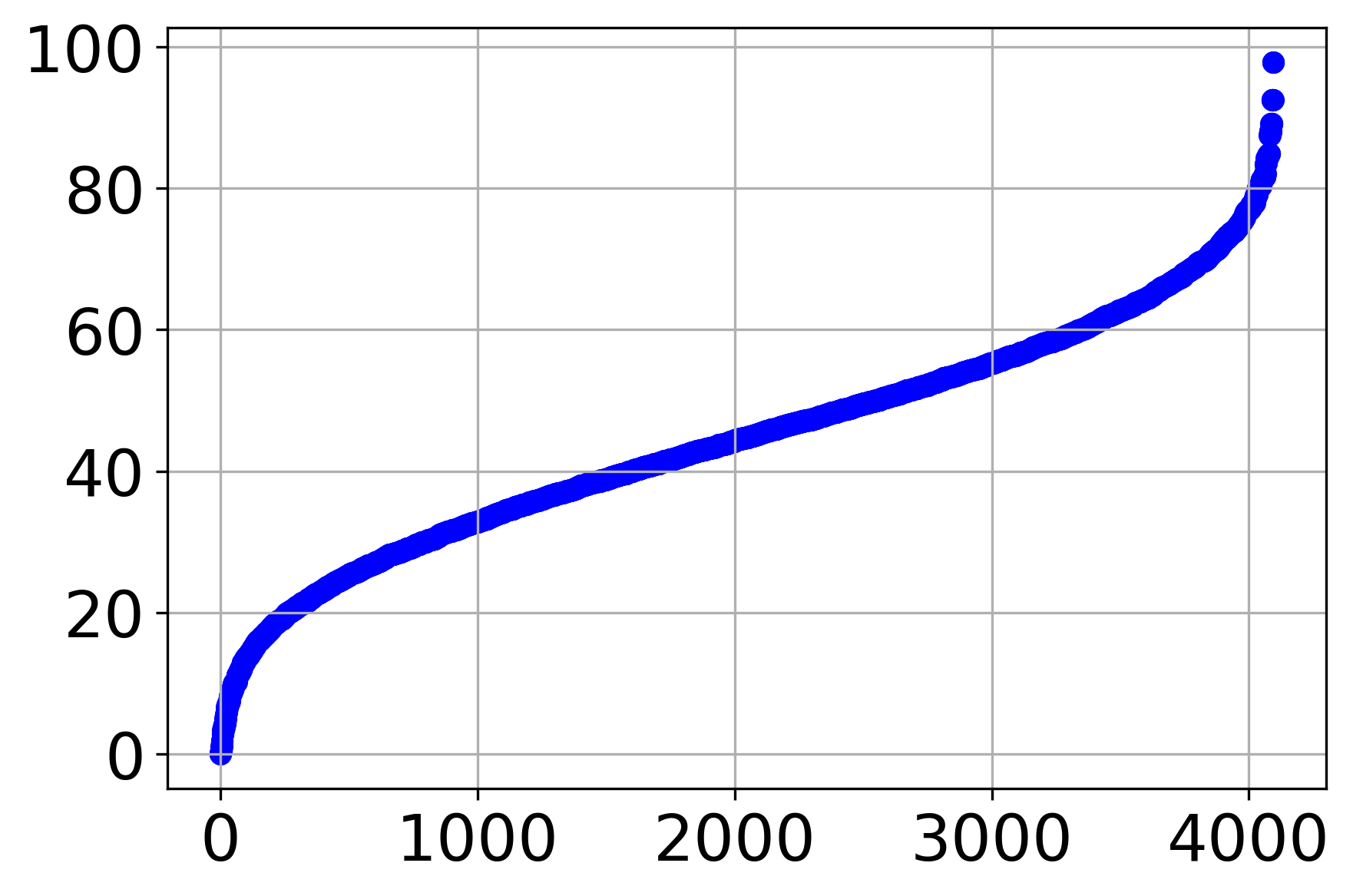}}}%
        %\put(10,620){(1)}%
        %\put(560,360){(2)}%
    \end{overpic}
    \caption{Spectrum of four-site linear chain with $A\neq 0$ and $J_2=0$.\label{SpectrumNoJ2}}
\end{figure}

Figure \ref{j2_pert_noA} shows the perturbative effect of $J_2$ on the system with only $J_1$ present (as in Fig.\ \ref{SpectrumNoSIANoJ2}). We see that the energy of each degenerate manifold increases linearly with $J_2$ at slightly different rates, but producing no splittings. Although one expects that a large enough $J_2$ would alter the polarization of the ground state, this does not happen for the small value of $J_2$ in the main text. 

\begin{figure}[]
    \begin{overpic}[width=8.6cm]{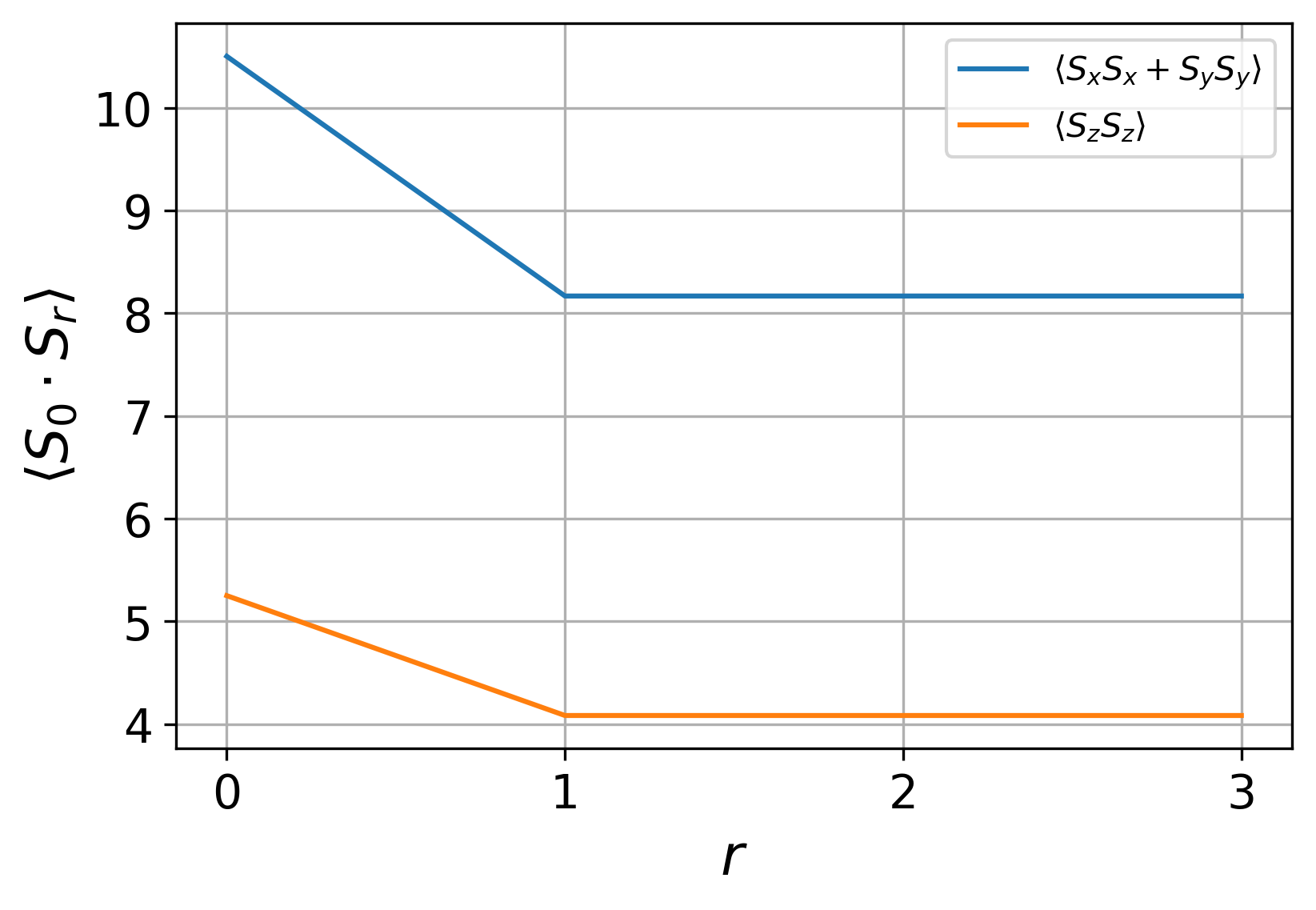}%
    \end{overpic}
    \caption{Spin-spin correlations for four-site chain for $J_2=A=0$. $r$ labels relative `site separation', where 0 is the first site of the chain.\label{L4-correlations-onlyJ1}}
\end{figure}

Figure \ref{L4-correlations-onlyJ1} shows that when only $J_1$ is present, the ground state has indeed isotropic ferromagnetic correlations, as one would expect.

\subsection{Role of single ion anisotropy}
\begin{figure}[]
    \begin{overpic}[width=8.6cm]{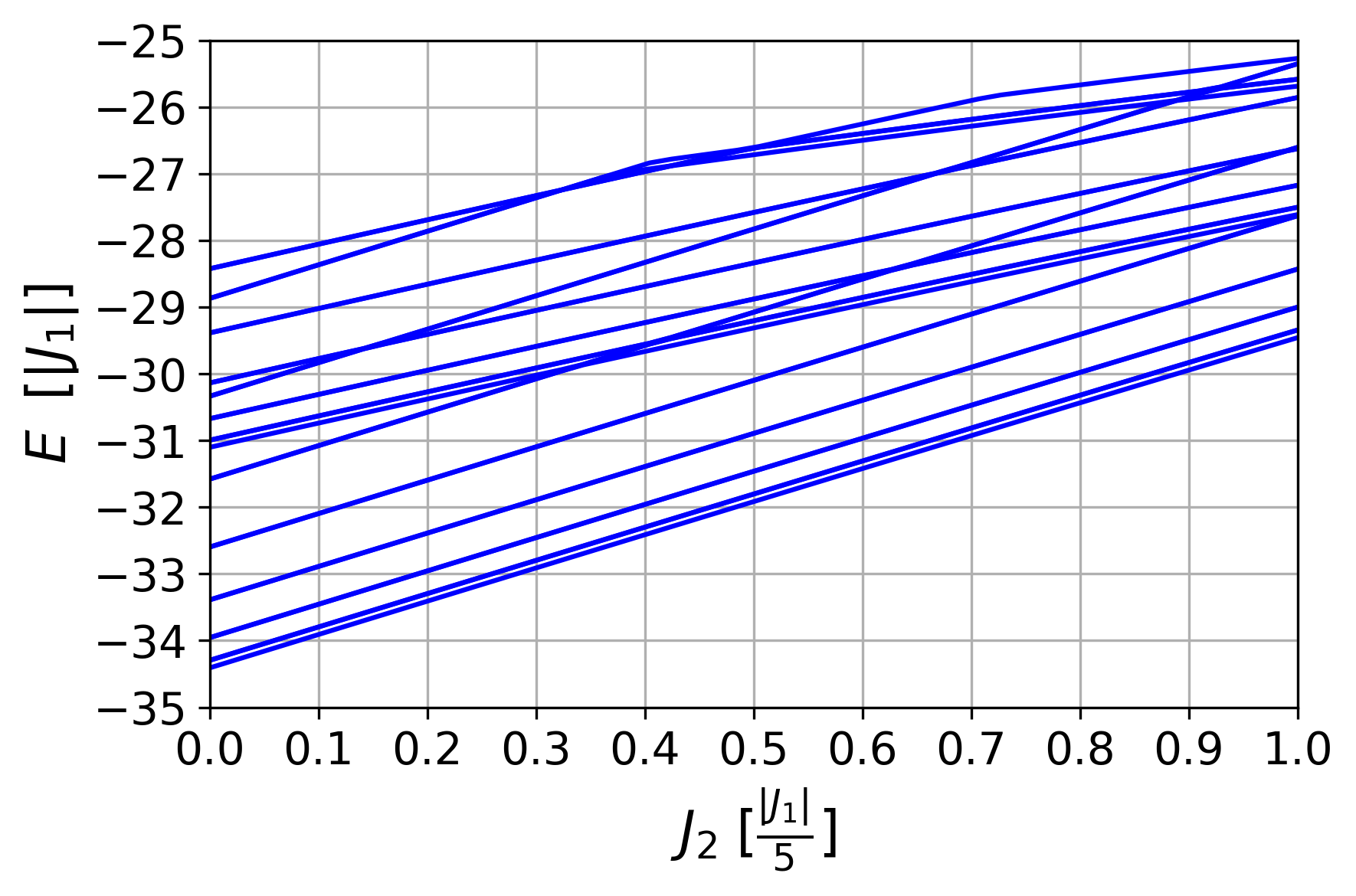}%
    \end{overpic}
    \caption{Spectrum of four-site linear chain with increasing $J_2$ antiferromagnetic coupling, while $J_1$ and SIA $A$ remain fixed at values in main text. The ground state remains non-degenerate, with linear energy increase with $J_2$ over this range.\label{FigY}}
\end{figure}

A similar behavior of monotonically increasing energy occurs when the magnetic anisotropy is considered.  The presence of $A\neq 0$ splits the degenerate ground state manifold, as seen in Fig.\ \ref{SpectrumNoJ2}.  However, increasing $J_2$ shifts the states without further doublet-splitting, as can be seen in Fig.\ \ref{FigY}.  A finite $J_2$ value, as in the main text, results in Fig.\ 2, qualitatively similar to Fig.\ \ref{SpectrumNoJ2}.

\begin{figure}[]
    \begin{overpic}[width=8.6cm]{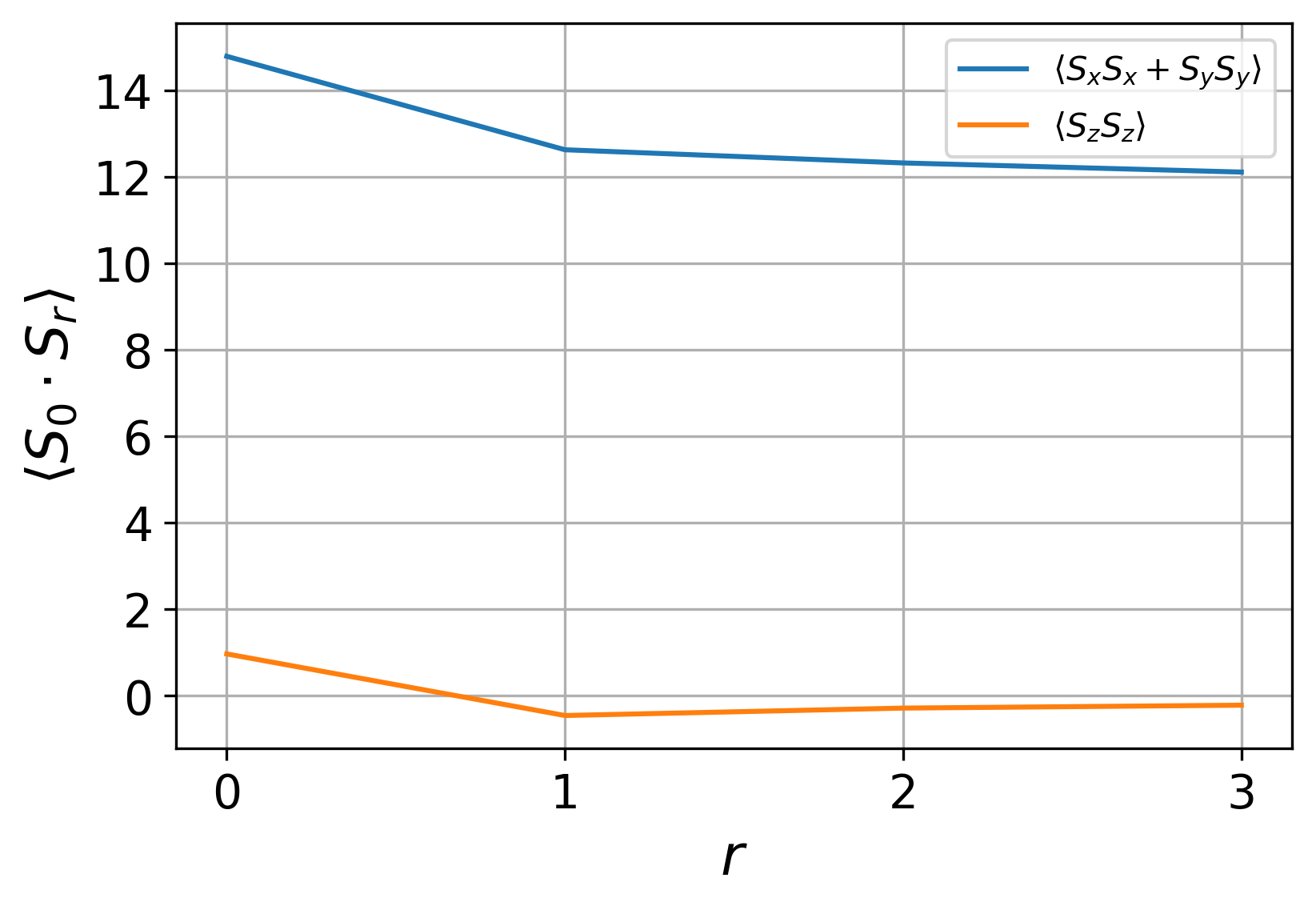}%
    \end{overpic}
    \caption{Different components of spin-spin correlations for $J_1,A \neq 0$. $r$ labels the relative `site separation'.\label{spin-spin-noJ2}}
\end{figure}

The presence of SIA ($A\neq 0$) splits the highly-degenerate ground state, as mentioned, but keeps the strong ferromagnetic correlations between sites.  As $A>0$, the `easy plane' ferromagnetic correlations exist only for in-plane components, shown in Fig.\ \ref{spin-spin-noJ2}, as expected.

\begin{figure}[]
    \begin{overpic}[width=8.6cm]{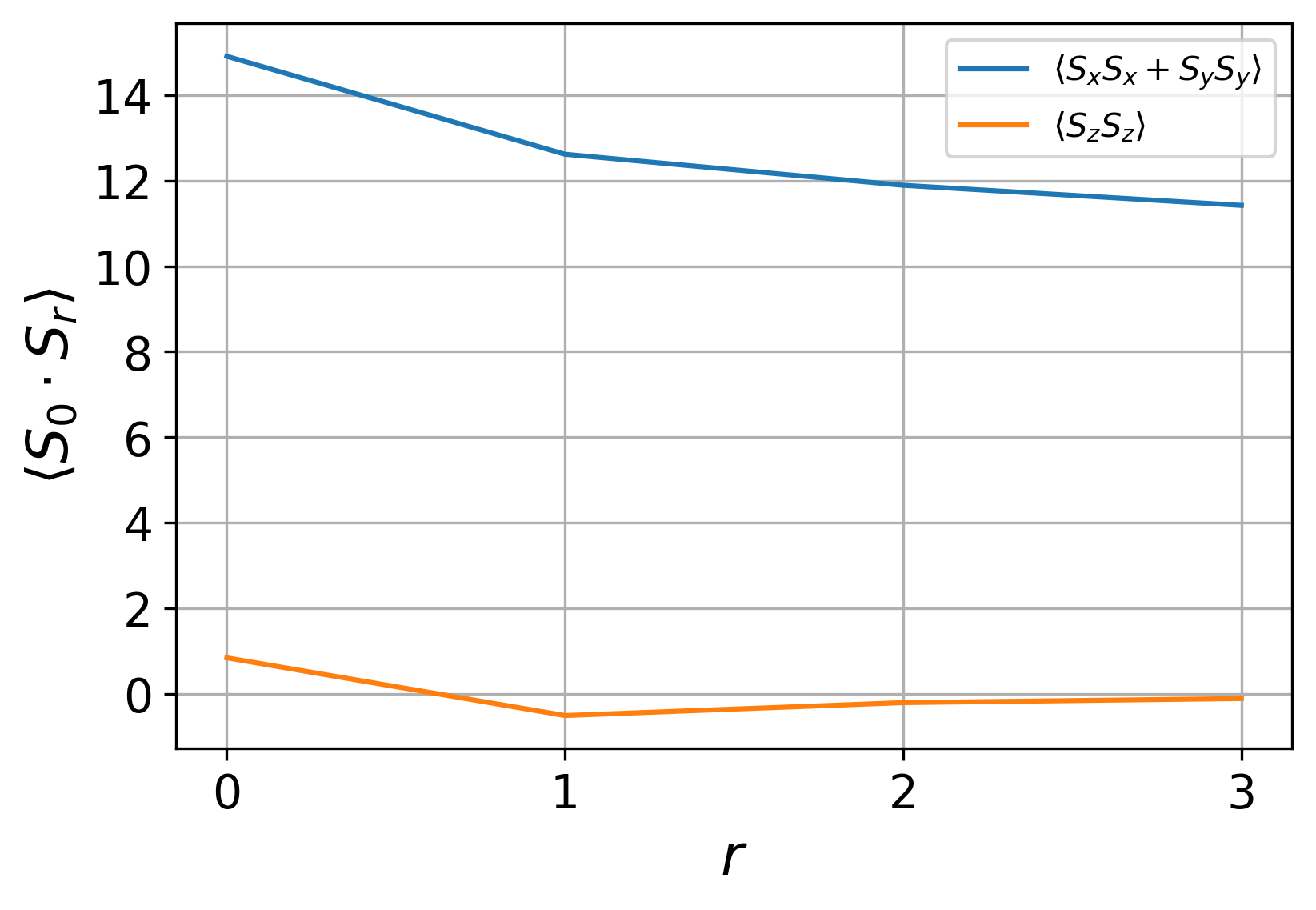}%
    \end{overpic}
    \caption{Different components of spin-spin correlations with all interactions contributing. Notice $r$ label denotes relative spin `separation'.\label{spin-spin}}
\end{figure}

When one considers all the interactions present in the Eu-on-gold system, i.e., $J_1$, $J_2$, and $A$ non-zero, the spin correlations in the ground state, Fig.\ \ref{spin-spin}, remain ferromagnetic in-plane, with only a slight decrease for the more-distant sites ($r=3$), as the effect of $J_2$ begins to suppress the ferromagnetic order.

\subsection{Successive spin flips and differential conductance}
The weight profiles described in the main text are shown for a single tunneling event (or `flip').  If one considers independent successive raising, lowering, or scattering events on the ground state, one
obtains a qualitative differential conductance map.  Notice that this does not consider relaxation events or interference between different processes, and ignores the likely weaker signal accompanying larger spin events--see discussion in main text.  

\begin{figure}[]
    \begin{overpic}[width=8.6cm]{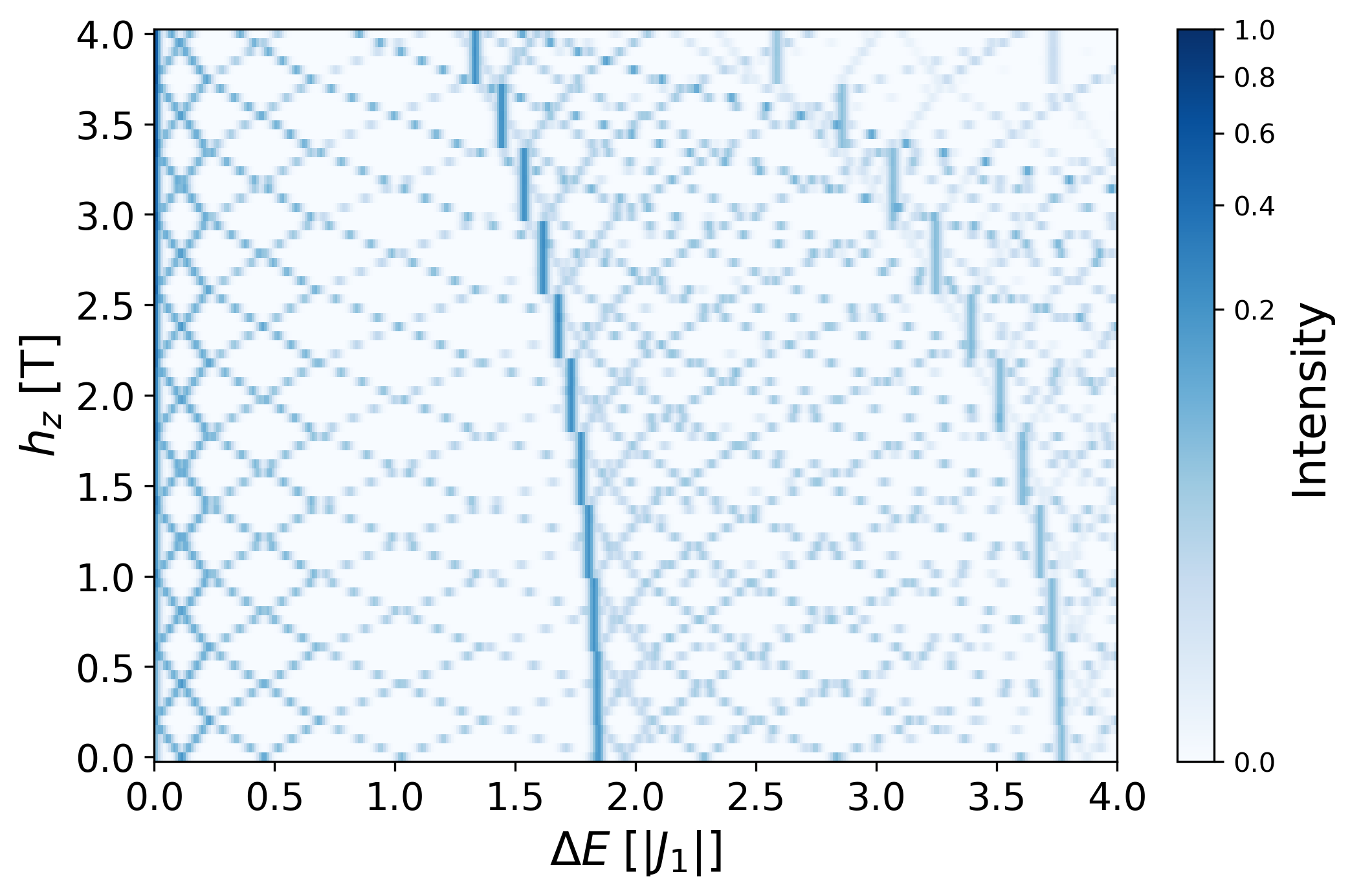}%
    \end{overpic}
    \caption{Heatmap of first site full tunneling.\label{first_site_full_tunneling}}
\end{figure}

With these caveats, Fig.\ \ref{first_site_full_tunneling} shows the action of the terms $(S^+)^n + (S^-)^n + (S^z)^n$, on the first site of the chain in its ground state. 
We emphasize that it is likely that higher $n$ contributions would be much weaker than the lower few. In other words, the latter would dominate the experimental differential conductance trace, as displayed in the main text. The low energy traces are essentially identical to those in the main text, although a much richer structure exists at higher energy and higher fields.

